\documentclass[twocolumn]{aastex631}  

\usepackage{graphicx}

\newcommand{\quotes}[1]{`#1'}

\shorttitle{ZTF QSO Catalog}
\shortauthors{Nakoneczny et al.}

\begin{document}

\title{
    QZO: A Catalog of 5 Million Quasars from the Zwicky Transient Facility\\
}

\correspondingauthor{Szymon~J.~Nakoneczny}
\email{nakoneczny@cft.edu.pl}

\author[0000-0003-2130-7143]{S.~J.~Nakoneczny}
\affiliation{Division of Physics, Mathematics and Astronomy, California Institute of Technology, Pasadena, CA 91125, USA}

\author[0000-0002-3168-0139]{M.~J.~Graham}
\affiliation{Division of Physics, Mathematics and Astronomy, California Institute of Technology, Pasadena, CA 91125, USA}

\author[0000-0003-2686-9241]{D.~Stern}
\affiliation{Jet Propulsion Laboratory, California Institute of Technology, 4800 Oak Grove Drive, Pasadena, CA 91109}

\author[0000-0003-3367-3415]{G.~Helou}
\affiliation{Division of Physics, Mathematics and Astronomy, California Institute of Technology, Pasadena, CA 91125, USA}

\author[0000-0002-0603-3087]{S.~G.~Djorgovski}
\affiliation{Division of Physics, Mathematics and Astronomy, California Institute of Technology, Pasadena, CA 91125, USA}

\author[0000-0001-8018-5348]{E.~C.~Bellm}
\affiliation{DIRAC Institute, Department of Astronomy, University of Washington, Seattle, WA 98195, USA}

\author[0000-0001-9152-6224]{T.~X.~Chen}
\affiliation{IPAC, California Institute of Technology, Pasadena, CA 91125, USA}

\author[0000-0002-5884-7867]{R.~Dekany}
\affiliation{Caltech Optical Observatories, California Institute of Technology, Pasadena, CA 91125, USA}

\author[0000-0003-0228-6594]{A.~Drake}
\affiliation{Division of Physics, Mathematics and Astronomy, California Institute of Technology, Pasadena, CA 91125, USA}

\author[0000-0003-2242-0244]{A.~A.~Mahabal}
\affiliation{Division of Physics, Mathematics and Astronomy, California Institute of Technology, Pasadena, CA 91125, USA}
\affiliation{Center for Data Driven Discovery, California Institute of Technology, Pasadena, CA 91125, USA}

\author[0000-0002-8850-3627]{T.~A.~Prince}
\affiliation{Division of Physics, Mathematics and Astronomy, California Institute of Technology, Pasadena, CA 91125, USA}

\author[0000-0002-0387-370X]{R.~Riddle}
\affiliation{Caltech Optical Observatories, California Institute of Technology, Pasadena, CA 91125, USA}

\author[0000-0001-7648-4142]{B.~Rusholme}
\affiliation{IPAC, California Institute of Technology, Pasadena, CA 91125, USA}

\author{N.~Sravan}
\affiliation{Department of Physics, Drexel University, Philadelphia, PA 19104, USA}

\begin{abstract}

Machine learning methods are well established in the classification of quasars (QSOs). However, the advent of light curve observations adds a great amount of complexity to the problem. Our goal is to use the Zwicky Transient Facility (ZTF) to create a catalog of QSOs. We process the ZTF DR20 light curves with a transformer artificial neural network and combine different surveys with extreme gradient boosting. Based on ZTF \textit{g}-band and WISE observations, we find 4,849,574 objects classified as QSOs with confidence higher than 90\%. We robustly classify objects fainter than the $5\sigma$ SNR limit at $g=20.8$ by requiring $g < n_\mathrm{obs} / 80 + 20.375$. For 33\% of QZO objects, with available WISE data, we publish redshifts with estimated error $\Delta z/(1 + z) = 0.14$. We find that ZTF classification is superior to the Pan-STARRS static bands, and on par with WISE and Gaia measurements, but the light curves provide the most important features for QSO classification in the ZTF dataset. Using ZTF \textit{g}-band data with at least 100 observational epochs per light curve, we obtain 97\% F1 score for QSOs. We find that with 3 day median cadence, a survey time span of at least 900 days is required to achieve 90\% QSO F1 score. However, one can obtain the same score with a survey time span of 1800 days and the median cadence prolonged to 12 days. We release the catalog, models and code on Zenodo\footnote{\url{https://zenodo.org/records/16410988}}\footnote{\url{https://zenodo.org/records/16535608}}, and GitHub\footnote{\url{https://github.com/snakoneczny/ztf-agn}}.

\end{abstract}


\section{Introduction} \label{sec:introduction}

Quasars are a type of active galaxy, usually occupying massive dark matter halos \citep{Eftekharzadeh_2015, DiPompeo_2016}, and emitting enormous amounts of energy through accretion of matter onto a supermassive black hole \citep{Kormendy_2013}. They mostly appear as faint and point like objects. Due to their highly biased tracing of vast volumes of the large scale structure \citep{DiPompeo_2014, Laurent_2017}, quasars are used for various cosmological applications: a test of the cosmological principle of isotropy and homogeneity \citep{Secrest_2021, Dam_2023}, the growth rate of structure \citep{Garcia_2021, Alonso_2023}, primordial non-Gaussianity \citep{Leistedt_2014, Castorina_2019, Krolewski_2024}, the Hubble distance \citep{Hou_2020}, baryon acoustic oscillations \citep{Ata_2017, Zarrouk_2021}, the integrated Sachs-Wolfe effect \citep{Stolzner_2018}, the expansion rate of the universe as standardizable candles \citep{Setti:1973, Risaliti_2015, Lusso_2020}, calibration of the reference frames for Galactic studies \citep{Lindegren_2018}, measurement of halo masses \citep{DiPompeo_2017}, cross-correlation with the $\gamma$-ray background \citep{Cuoco_2017}, tomographic angular clustering \citep{Leistedt_2014, Ho_2015}, quasar bias measurements \citep{Sherwin_2012}, the baryon density \citep{Yahata_2005}, cosmic magnification \citep{Scranton_2005}, and cross correlation with foreground galaxies as a probe of weak lensing \citep{Menard_2002}.

Quasars are also variable objects, which can be described by a damped random walk \citep{Kelly_2009, Kozlowski_2010}. The model is consistent with the observed light curves on timescales from several months to a few years \citep{MacLeod_2010, Zu_2011, MacLeod_2012}, but diverges at scales less than a few months \citep{Mushotzky_2011, Zu_2013, Kasliwal_2015, Stone_2022}. The physical source of quasar variability is not well understood \citep{Ulrich_1997, Padovani_2017}. Studies suggested that variability amplitude increases with decreasing luminosity, rest frame wavelength, and Eddington ratio \citep{Wills_1993, Giveon_1999, Berk_2004, Wold_2007}, and the characteristic variability time-scale depends on the black hole mass \citep{Collier_2001, Kelly_2009, MacLeod_2010, Simm_2016}. \cite{Burke_2021} found a strong correlation between the damping time-scale and black hole mass, and \citep{Stone_2022} found a weak wavelength dependence of damping time scale.

There are also fluctuations in a form of, e.g. tidal disruption events \citep[TDEs, ][]{Rees_1988, Gezari_2021, Stein_2024}, and possibly gravitational wave (GW) flares \citep{McKernan_2019, Graham_2020, Wang_2021, Kimura_2021, Perna_2021}. The latter have not yet been confirmed, but \cite{Graham_2023} present nine candidate counterparts for such events from Ligo/Virgo detectors \citep{Ligo_2015, Acernese_2015}. The importance of such events results from accretion disks being the only channel for stellar mass black hole mergers which reliably produce electromagnetic (EM) counterparts, and thus the identification of GW event with a host galaxy at a known redshift \citep{Ashton_2021, Calderon_2021, Palmese_2021}. The GW signal then becomes a standard siren, which can be used to measure the expansion history of the universe \citep{Mukherjee_2020, Chen_2022}. However, detection of EM counterparts in AGN (active galactic nucleus) time series is challenging due to potentially too bright or too thick accretion disks, as well as the incomplete census of AGN. Hence, a first step in a systematic search of such flares is a construction of AGN catalog with available time-series data.

The most reliable source of quasar\footnote{We use AGN and QSO terms interchangeably.} identification and redshifts are spectra. Spectroscopic surveys have provided $\sim 10^5$-$10^6$ QSOs, e.g. 2dF QSO Redshift Survey \citep[2QZ,][]{Croom_2004}, 2dF-SDSS LRG and QSO \citep[2SLAQ,][]{Croom_2009}, and the Sloan Digital Sky Survey \citep[SDSS,][]{York_2000, Lyke_2020}. Ongoing and future spectroscopic surveys such as DESI \citep{DESI_2016} and 4MOST \citep{deJong_2019, Merloni_2019, Richard_2019} expect to obtain spectra for 3 million quasars. Since spectroscopic surveys are time-consuming, wide area photometric surveys are a much better way of observing large numbers of QSOs. However, the lack of spectra makes classification more complicated and redshifts much less accurate. Nearly 3 million quasars with photometric redshifts (photo-zs) have been cataloged \citep[e.g. ][]{Shu_2019, Kunsagi_2022, Yang_2023, Nakazono_2024, Storey-Fisher_2024}, through surveys such as the Wide-field Infrared Survey Explorer \citep[WISE, ][]{Wright_2010}, unWISE \citep{Lang_2014}, Pan-STARRS \citep{Chambers_2016}, and Gaia \citep{Gaia_2016}. In the future, the Rubin Observatory’s LSST will photometrically observe up to 10 million quasars \citep{Ivezic_2016}.

One possibility to classify objects in photometric data is SED fitting, which depending on available data can derive physical properties \citep{Ciesla_2015, Stalevski_2016, Calistro_2016, Yang_2020, Malek_2020}, as well as photo-zs \citep{Salvato_2009, Salvato_2011, Fotopoulou_2016, Fotopoulou_2018}. Also, using different magnitudes, one can create a color-color space which allows for object classification \citep{Warren_2000, Maddox_2008, Edelson_2012, Stern_2012, Wu_2012, Secrest_2015, Assef_2018}. More sophisticated methods include probabilistic models \citep{Richards_2004, Richards_2009a, Richards_2009b, Bovy_2011, Bovy_2012, DiPompeo_2015, Richards_2015}, but due to the high complexity of multidimensional color-color space, machine learning (ML) methods achieve the best results \citep{Brescia_2015, Carrasco_2015, Kurcz_2016, Nakoneczny_2019, Logan_2020}. One can also estimate photo-zs with ML \citep{Brescia_2013, Yang_2017, Pasquet_2018, Curran_2020, Nakoneczny_2021}.

Another useful information for quasar classification is the photometric variability across time, provided by surveys such as ASAS \citep{Pojmanski_2002}, NSVS \citep{Wozniak_2004, Hoffman_2009}, the Palomar Transient Factory \citep{Law_2009}, the Catalina Surveys \citep{Drake_2014, Drake_2017}, and ASAS-SN \citep{Kochanek_2017, Jayasinghe_2018}. The Zwicky Transient Facility \citep[ZTF, ][]{Bellm_2019, Graham_2019, Masci_2019, Dekany_2020} has operated since March 2018 and covers the northern sky with 3 day cadence. Since the observations of quasar light curves are relatively new, and their usage in classification has not yet been fully established in comparison to other astronomical observations, ZTF is well suited for this goal, as it delivers light curves with irregular sampling, over a six year period of observations.

Several projects in ZTF performed AGN classification alongside other classes of variable and non variable objects. For instance, the ZTF Source Classification Project \citep[SCoPe, ][]{Roestel_2021b, Coughlin_2021, Healy_2024} uses convolutional neural networks with time magnitude histograms, and extreme gradient boosting \citep[XGB, ][]{Chen_2016} with precomputed features, trained on a set of $\sim 10^3$ manually selected AGNs. Automatic Learning for the
Rapid Classification of Events \citep[ALeRCE, ][]{Forster_2021, Sanchez_2021} classifies the ZTF alert stream, which is a subset of ZTF light curves showing large fluctuation in magnitude, using balanced decision trees with precomputed features, and distinguishing several AGN subclasses. \citep{Sanchez_2023} classifies light curves within the overlap of ZTF and 4-m Multi-Object Spectroscopic Telescope \citep[4MOST, ][]{deJong_2019} surveys, distinguishing 3 ranges of redshifts for the AGNs, in order to identify AGN candidates for 4MOST survey. Our goal is to classify all ZTF light curves using a deep learning transformer model trained on SDSS data with 120,000 quasars in our training dataset.

We test quasar classification with deep learning methods based on the transformer architecture, and compare the results with shallow feature extraction methods. Next, we establish the minimum requirements on the length and sampling of time-series data in order to achieve reliable quasar classification. We compare ZTF classification to visible \textit{griz} bands from Pan-STARRS DR1 \citep[PS, ][]{Chambers_2016}, mid-infrared W[1-4] filters from AllWISE \citep{Wright_2010, Cutri_2014}, and parallax and proper motion measurements from Gaia EDR3 \citep{Gaia_2016, Gaia_2021}. Finally, we use these results to create a high quality quasar catalog based on the ZTF data.

The paper is structured as follow. Section \ref{sec:data} presents the ZTF survey, as well as inference and training data for the ML models. Section \ref{sec:methodology} presents the ML models, evaluation methods, and experimental methodology. Section \ref{sec:results} presents the results, while Section \ref{sec:observations} briefly discusses spectroscopy obtained at Palomar Observatory of sources with conflicting SDSS and machine learning classifications. Finally, Section \ref{sec:conclusion} interprets and concludes the work. Unless stated otherwise, we present the results for the ZTF \textit{g}-band data.

\section{Data} \label{sec:data}

\subsection{Zwicky Transient Facility}\label{sec:ztf}

ZTF is an optical time domain survey operated at the Palomar 48 inch Schmidt telescope since March 2018. Typical limiting magnitudes in \textit{g} and \textit{r} bands are 20.8 and 20.6, respectively. During its initial phase, its public surveys observed 27,500 square degrees over the northern sky, with a three night cadence, and 1000 to 2000 square degrees at the Galactic plane, with a one night cadence \citep{Bellm_2019_b}. Since December 2020, the ZTF second phase, the northern sky has been observed every two nights. Additional private surveys have sampled the sky with more specialized cadences, such as continuous single-field ``deep drilling'' observations. The data processing and access is available through the Science Data System at the Infrared Processing and Analysis Center \citep[IPAC, ][]{Masci_2019}. The primary science goals of ZTF are the physics of supernovae and relativistic explosions, multi-messenger astrophysics, supernova cosmology, active galactic nuclei, TDEs, stellar variability, and solar system objects \citep{Graham_2019}.

\subsection{Inference data}\label{sec:inference_data}

We perform the inference using ZTF data release (DR) 20, covering March 2018 to January 2024. For inference, we use only the \textit{g}-band data, while we add the \textit{r}-band to experiments, and omit the \textit{i}-band due to its much lower coverage compared to the two other bands. In total, ZTF DR20 consists of 1.5B and 2.4B objects in the \textit{g}- and \textit{r}-bands, respectively. The estimated number of bad epochs per light curve, with likely suspect and unusable photometry, is 8\% and 11\% in \textit{g} and \textit{r}, respectively\footnote{See Table 1 in \url{https://irsa.ipac.caltech.edu/data/ZTF/docs/releases/dr20/ztf_release_notes_dr20.pdf}}. We remove deep drilling observations by averaging observations within the same night. Finally, after removal of light curves with less than 20 observations, we obtain the inference dataset of 789M objects in the ZTF \textit{g} band. Additionally, we either mark or remove duplicated light curves by finding objects which within $1\arcsec$ have a neighbor with more observation epochs. The number of deduplicated light curves equals 533M. In this work, we do not consider morphological features, but these can be used as an extension to our methodology.

\subsection{Training data}\label{sec:training_data}

\begin{figure*}
    \gridline{
      \fig{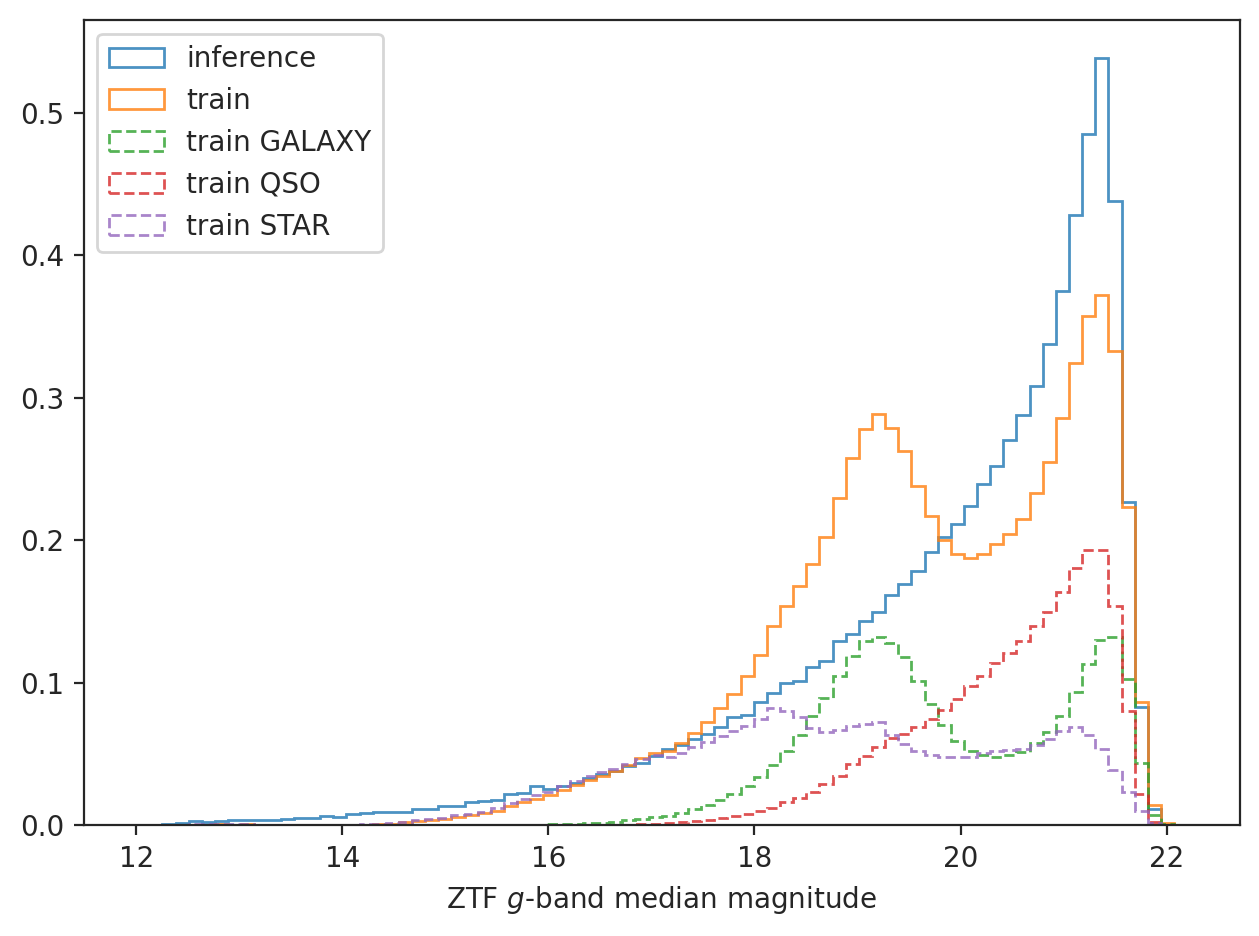}{0.5\textwidth}{}
      \fig{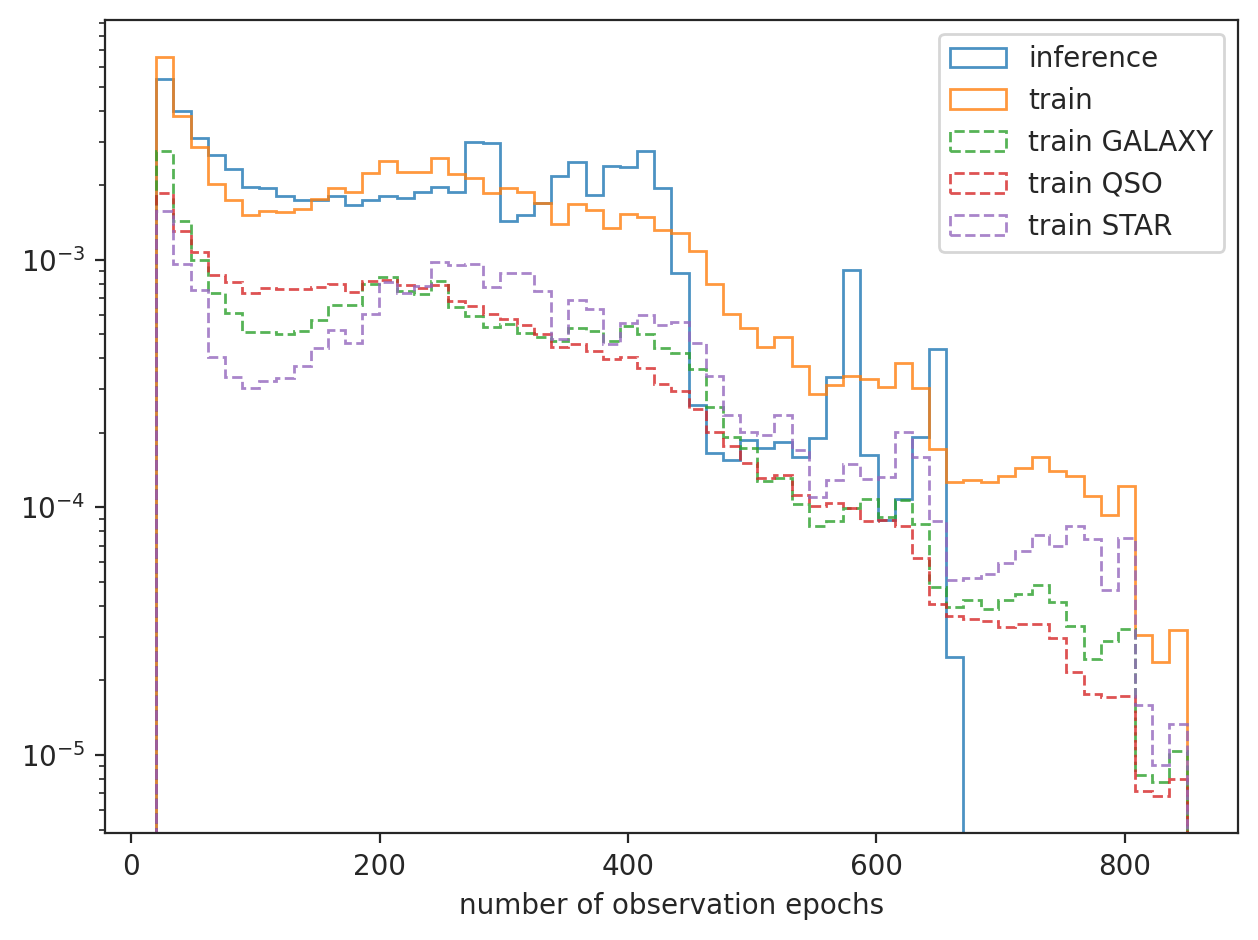}{0.5\textwidth}{}
    }
    \caption{Distributions of median \textit{g}-band magnitude and number of observation epochs for the inference (blue), training (orange), and each class component of the training data (green, red, and purple). The inference and training distributions are normalized to unity, and the class distributions are normalized to one third.}
    \label{fig:data}
\end{figure*}

We obtain the training data in \textit{g} and \textit{r} bands by cross-matching the ZTF data with SDSS DR18 \citep[SDSS,][]{Almeida_2023}, using a $1\arcsec$ distance threshold, and applying the same preprocessing as in the case of inference data. We clean the SDSS data by removing objects with known redshift estimation errors as encoded in the zWarning flag, with the exception of the fifth bit of this flag, which does not signal problems. Additionally, we remove duplicated ZTF light curves by accepting only the longest light curve among the neighbors within the $1\arcsec$ distance. We analyze both \textit{g}- and \textit{r}-band datasets independently. The training set size is 483k and 665k in \textit{g}- and \textit{r}-bands. The class distribution is 47\% galaxies, 25\% quasars, and 28\% stars in the \textit{g}-band, and 59\% galaxies, 19\% quasars, and 22\% stars in the \textit{r}-band. It totals to 645062 and 703685 quasars in \textit{g}- and \textit{r}-bands, respectively. In Section \ref{sec:lc_classification}, we compute the experiments with shallow precomputed features \citep[see Table 1 in][]{Healy_2024}. For this goal, we use ZTF DR5 published in January 2021, since at the time of writing, shallow features were only computed for this DR. However, we did not first utilize all the light curves from this dataset, and could not update it later due to removal of data from the servers. Hence, the DR5 training dataset size is 301k and 381k in \textit{g}- and \textit{r}-bands, respectively, which also lowers the deep learning results due to a smaller training set size. We use the DR5 data only for comparison between the shallow and deep learning results of light curve classification. We use 37 of the precomputed features, omitting the magnitude-time histograms and ZTF alerts information.

Figure \ref{fig:data} shows distributions of median magnitude and number of observations for inference and training data, also separated for different classes. The plots are normalized to unity, and the class plots are divided by three in order to properly compare the different classes. We can see a bimodal galaxy magnitude distribution, which does not bias the final inference, as we show in Fig. \ref{fig:mag_inference}. The training data cover the faint end of the inference data. However, the training distribution is different from the inference one, which together with the class distribution, hint to a larger fraction of quasars present at the faint end of inference data. Due to this, the final classification of inference data requires additional testing, and the results will not be calibrated. It means that if we chose 100 objects from the QSO classification probability range 90\% to 91\%, the actual number of QSOs in such a sample would be different than 90 or 91 objects, and that would not be due to a chance occurrence. The number of observations plot shows galaxies, quasars, and stars within the whole range of epochs. A shift in these distributions could also bias the final results.

\subsection{Other surveys}\label{sec:cross_matches}

\begin{figure}
    \fig{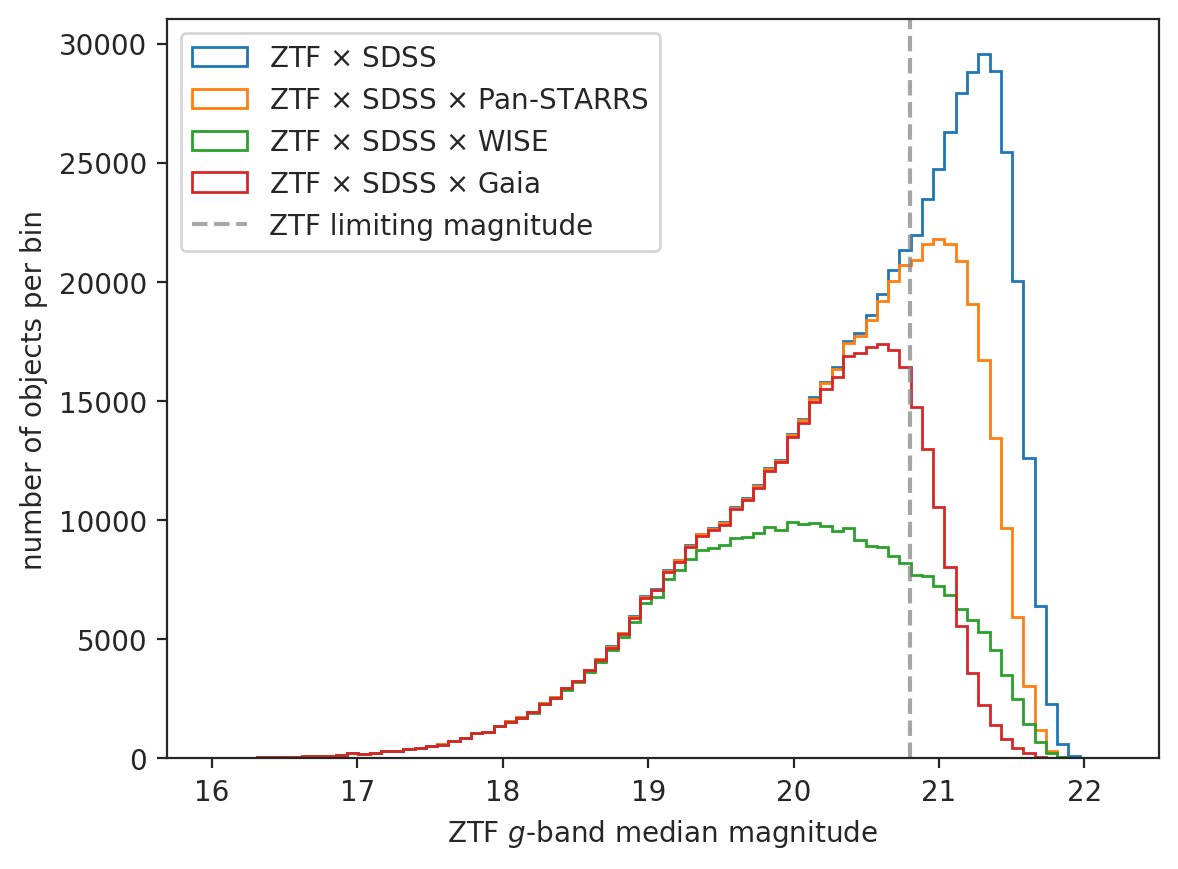}{0.5\textwidth}{}
    \caption{Number of quasars available in the training sample (blue), and its cross-matches with PS (orange), WISE (green) and Gaia (red). The vertical line shows the ZTF's magnitude limit, while for the other surveys, we apply limiting magnitude cuts as described in Section \ref{sec:cross_matches}. We note that the final catalog has no magnitude cuts on WISE survey.}
    \label{fig:surveys}
\end{figure}

We compare the ZTF based classification with AllWISE, Gaia EDR3, and PS DR1 by cross-matching the ZTF training datasets in \textit{g}- and \textit{r}-bands using a $1\arcsec$ matching distance. From PS, AllWISE, and Gaia, we use \textit{griz}, W[1-4], and \textit{g}, \textit{bp}, \textit{rp} magnitudes, respectively. We add colors from all magnitude pair combinations within each survey. Additionally, we add parallax and proper motion measurements from Gaia. We do not use error measurements for any of the features, either as inputs or weights for loss functions. Requiring availability of all the cross-matched features, we obtain the following subsets of the ZTF \textit{g}-band training data: ZTF x PS 98\%, ZTF x WISE 77\%, ZTF x Gaia 45\%. When comparing the classification based on different surveys, we apply the magnitude limits: PS \textit{griz} 22.0, 21.8, 21.5, 20.9 AB, WISE W[1-2] 17.1, 15.7 Vega, Gaia \textit{g} 21 Vega, respectively. Fig. \ref{fig:surveys} shows a comparison of quasar distributions between the full training sample and the cross-matched subsets. We also cross-match with AllWISE for the inference data, as justified in Sections \ref{sec:other_surveys} and \ref{sec:redshift}. For simplicity, we use WISE to refer to the AllWISE sample. When comparing ZTF with the other surveys, we require the features from cross-matched surveys to be present. However, for the final catalog, we add AllWISE features and process the missing magnitudes. During training, XGB marks each split in the trees as more suitable for the objects with missing features, which allows to later classify such objects.

\section{Methodology} \label{sec:methodology}

\subsection{Astromer}\label{sec:astromer}

We classify the light curves using Astromer \citep{Donoso_2023}, a deep learning transformer architecture, pre-trained on all the ZTF DR10 \textit{g}-band data. The model processes time series of magnitude and observation date using attention layers trained in a semi supervised way by removing random windows from the light curves, and then recreating the missing parts. The model is limited to 200 input observation epochs, transforming each observation into a vector of 256 tokens, which results in an encoder output size of $200\times256$. We first retrain the model in its original self-supervised training procedure, but using the training data from ZTF DR20. Then, we adapt the model for a classification task by processing the encoder output with other types of layers and additionally fine tuning the encoder weights.

\subsection{Pipeline}\label{sec:pipeline}

We test the ML models with a random 80\%, 10\%, 10\% split for the training, validation and test sample, respectively. We always use the same split for fine tuning Astromer, training the classification network, and making ensembles with other surveys. We use two metrics to evaluate the classification: the general three class accuracy defined as a fraction of correctly classified objects, and QSO F1 score defined as a harmonic mean of QSO completeness and precision.

We first retrain Astromer using its original training procedure on the \textit{g}- and \textit{r}-band training data, which do not contain the deep drilling observations unlike the original Astromer training. For the \textit{r}-band, this step involves transfer learning from the \textit{g}-band. Hence, the retraining for \textit{r}-band is about 3 times longer. We find that retraining the transformer model provides better results for the final classification task, and allows for faster training and experimentation with the classification models. We then tested the retrained Astromer for classification with different combinations of convolutional, long short-term memory \citep[LSTM,][]{Hochreiter_1997}, normalization, and dense layers. We tested between 1 and 6 dense layers of size ranging from 2048 to 128 neurons. W also tested reduction of the transformer embeddings to mean, maximum or minimum values. We find that the best results are achieved with the following architecture: mean reduction of the embeddings, 1024, 512 and 256 neuron dense layers with the rectified linear unit activation function \citep[ReLu, ][]{Agarap_2018}, normalization layer, and three neuron output layer. For training, we use the Adam optimizer, and we tested learning rate values lower or equal to 0.1. We arrive at the best solution with a 0.001 learning rate, scheduled to 0.0005 at epoch 4, and 0.0001 at epoch 9. After training artificial neural networks (ANN) on \textit{g}- and \textit{r}-bands, we use XGB to ensemble the deep learning classifications with features from the PS, WISE and Gaia surveys. We use the same XGB training parameters as found optimal in a similar classification task in \cite{Nakoneczny_2021}. We also test adding the features from other surveys as additional inputs to fully connected layers of the ANN, but find this approach to provide worse results. Also, ensembling with XGB allows separation of the ANN training from the final classification models, enabling for easier and faster comparisons with other surveys.

\section{Results} \label{sec:results}

\subsection{Light curve classification} \label{sec:lc_classification}

\begin{figure*}
    \gridline{
      \fig{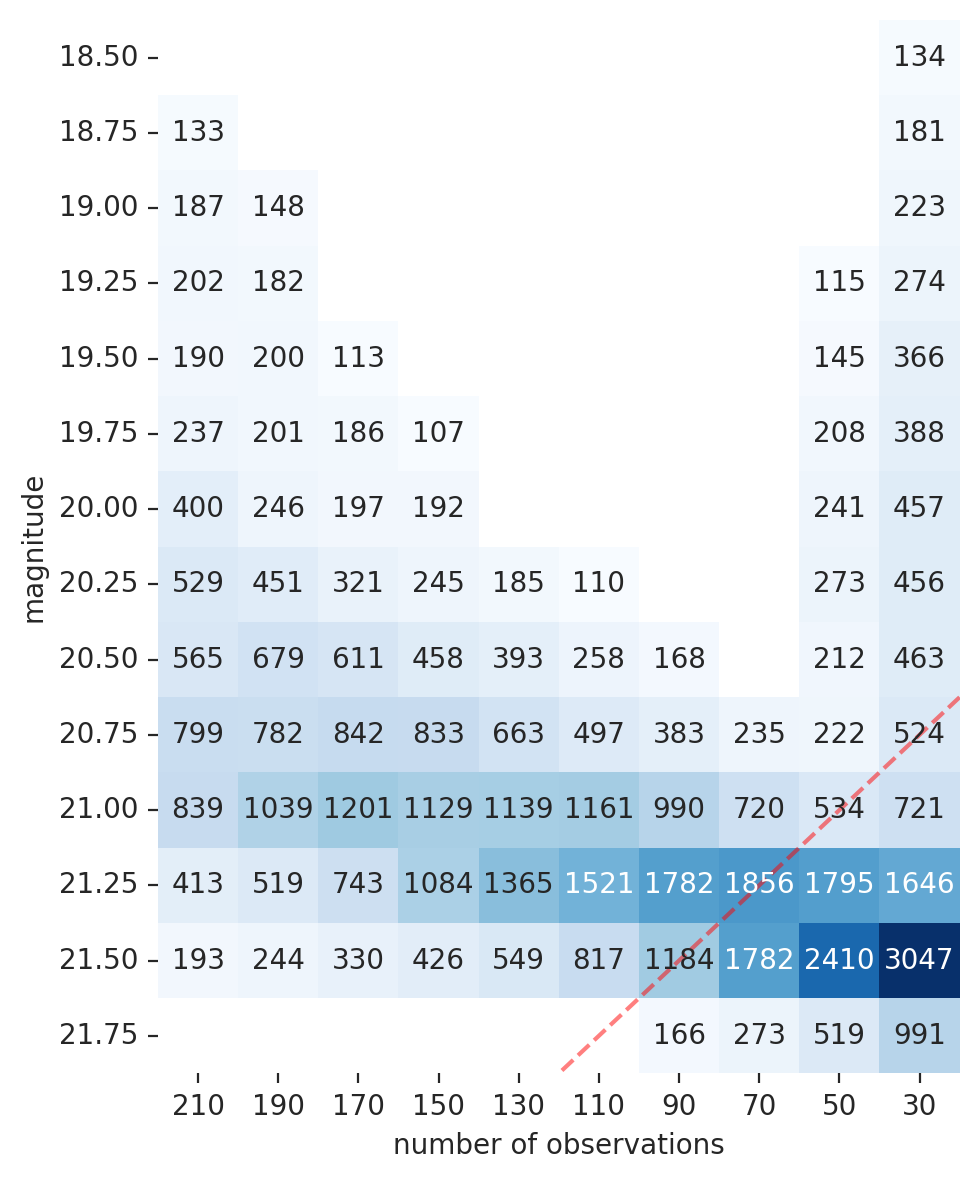}{0.5\textwidth}{}
      \fig{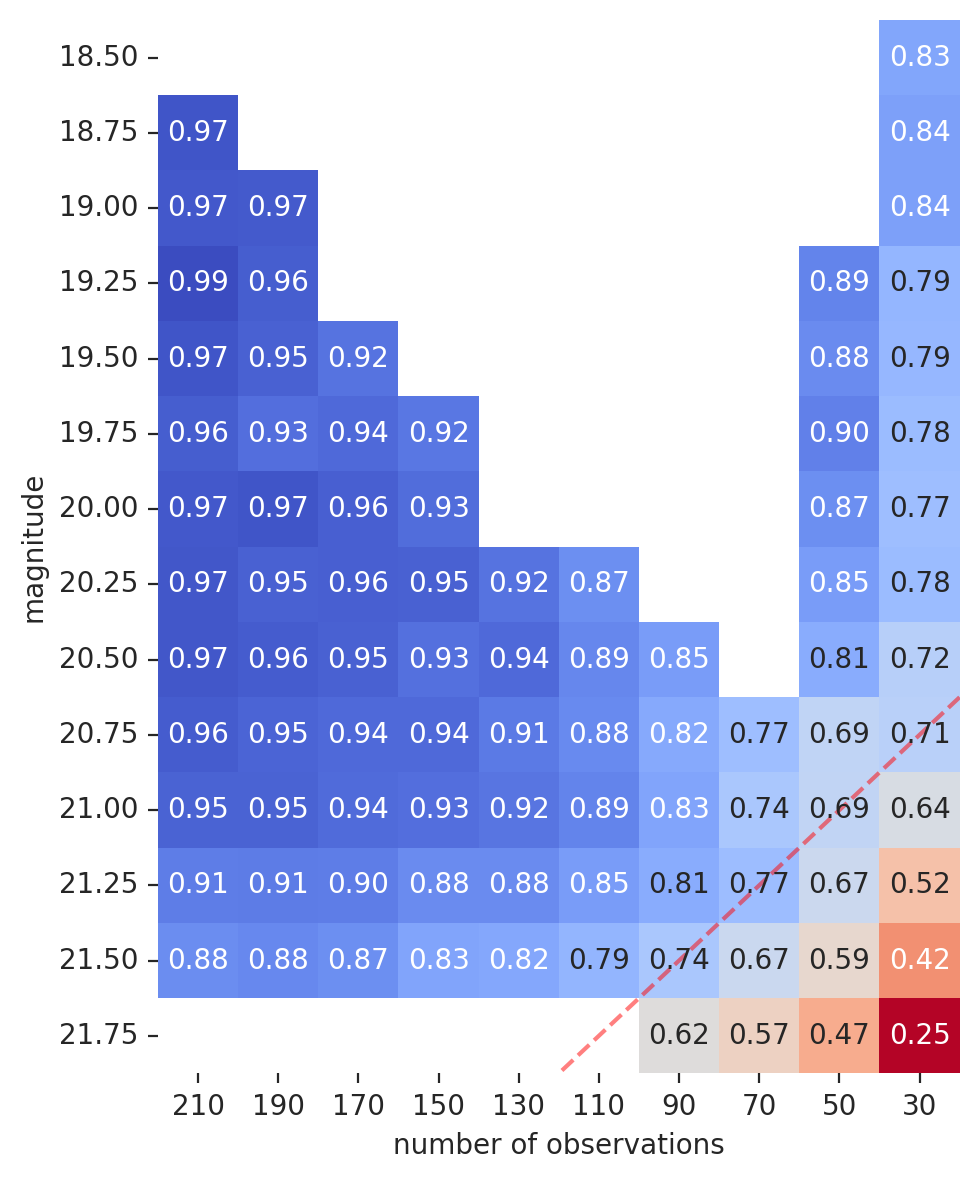}{0.5\textwidth}{}
    }
    \caption{Number of QSOs \textit{(left)} and QSO F1 scores \textit{(right)} for the ZTF \textit{g}-band XGB model based on transformer classifications and median magnitudes as features. We show the results in equally sized and non overlapping bins of the number of observations ($x$ axis) and \textit{g}-band magnitude ($y$ axis), where the ticks mark centers of the bins.}
    \label{fig:heat}
\end{figure*}

\begin{deluxetable*}{lcccccccc}
\tablenum{1}
\tablecaption{Classification results of the ZTF \textit{g}-band XGB model based on transformer classifications and median magnitudes as features, for combinations of limiting magnitude defined by mean SNR higher than $5\sigma$, ZTF observation epochs greater than 100, and a custom cut as a function of both magnitude and number of observation epochs.}
\tablewidth{0pt}
\tablehead{
\nocolhead{} & \multicolumn4c{\textit{g}-band} & \multicolumn4c{\textit{r}-band} \\
\nocolhead{} & \colhead{QSO} & \multicolumn2c{QSO F1} & \colhead{accuracy} & \colhead{QSO} & \multicolumn2c{QSO F1} & \colhead{accuracy} \\
\colhead{data cuts} & \colhead{fraction} & \colhead{full sample} & \colhead{subset} & \colhead{subset} & \colhead{fraction} & \colhead{full sample} & \colhead{subset} & \colhead{subset}
}
\decimalcolnumbers
\startdata
none & 1.00 & 0.88 & 0.88 & 0.88 & 1.00 & 0.85 & 0.85 & 0.89 \\
$g < n_\mathrm{obs} / 80 + 20.375$ & 0.84 & 0.85 & 0.93 & 0.91 & 0.90 & 0.84 & 0.89 & 0.90 \\ 
$\mathrm{SNR} > 5\sigma \lor n_\mathrm{obs} > 100$  & 0.79 & 0.83 & 0.94 & 0.92 & 0.84 & 0.82 & 0.90 & 0.91 \\
$n_\mathrm{obs} > 100$ & 0.73 & 0.80 & 0.95 & 0.93 & 0.76 & 0.79 & 0.92 & 0.92 \\
$\mathrm{SNR} > 5\sigma$ & 0.57 & 0.68 & 0.95 & 0.94 & 0.58 & 0.67 & 0.92 & 0.93 \\
$\mathrm{SNR} > 5\sigma \land n_\mathrm{obs} > 100$ & 0.50 & 0.64 & 0.97 & 0.95 & 0.50 & 0.63 & 0.95 & 0.94 \\
\enddata
\tablecomments{\quotes{QSO fraction} gives a fraction of quasars after the cuts with respect to the full training sample in the first row, \quotes{QSO F1 full sample} assumes negative classification for quasars excluded by the cuts, \quotes{QSO F1 subset} takes into the account only objects left after the data cut, \quotes{accuracy subset} stands for the general three class accuracy of subsets. Class distribution affects the accuracy score, in contrast to the F1 score, which can be applied to unbalanced class problems.}
\label{tab:cuts}
\end{deluxetable*}

\begin{figure*}
    \gridline{
      \fig{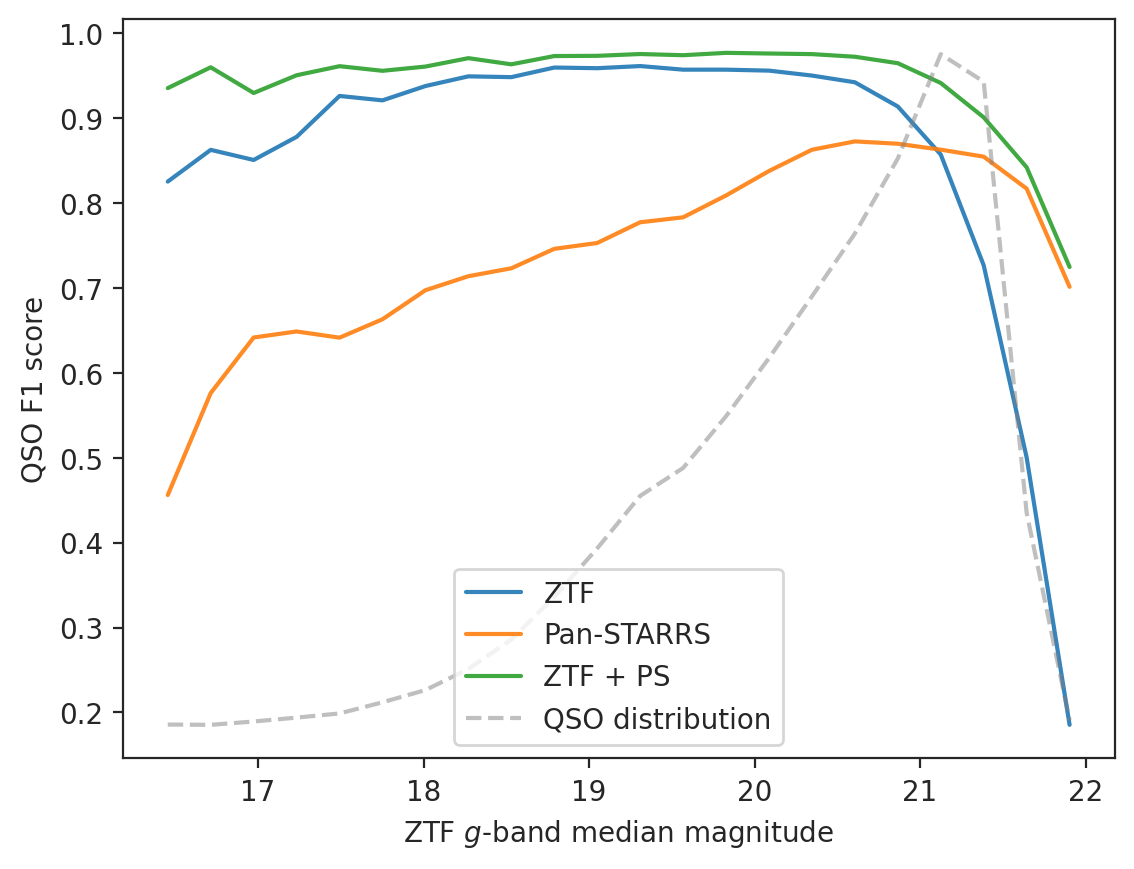}{0.5\textwidth}{}
      \fig{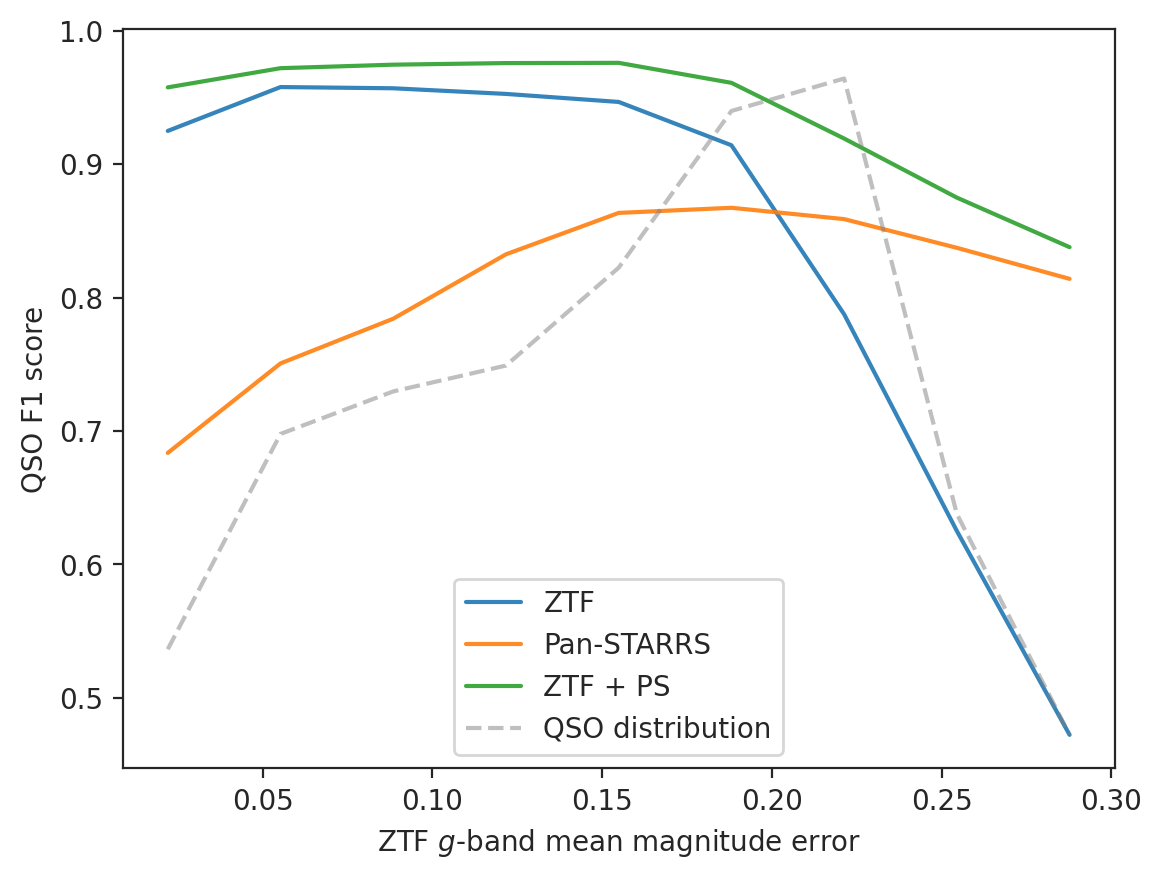}{0.5\textwidth}{}
    }
    \gridline{
      \fig{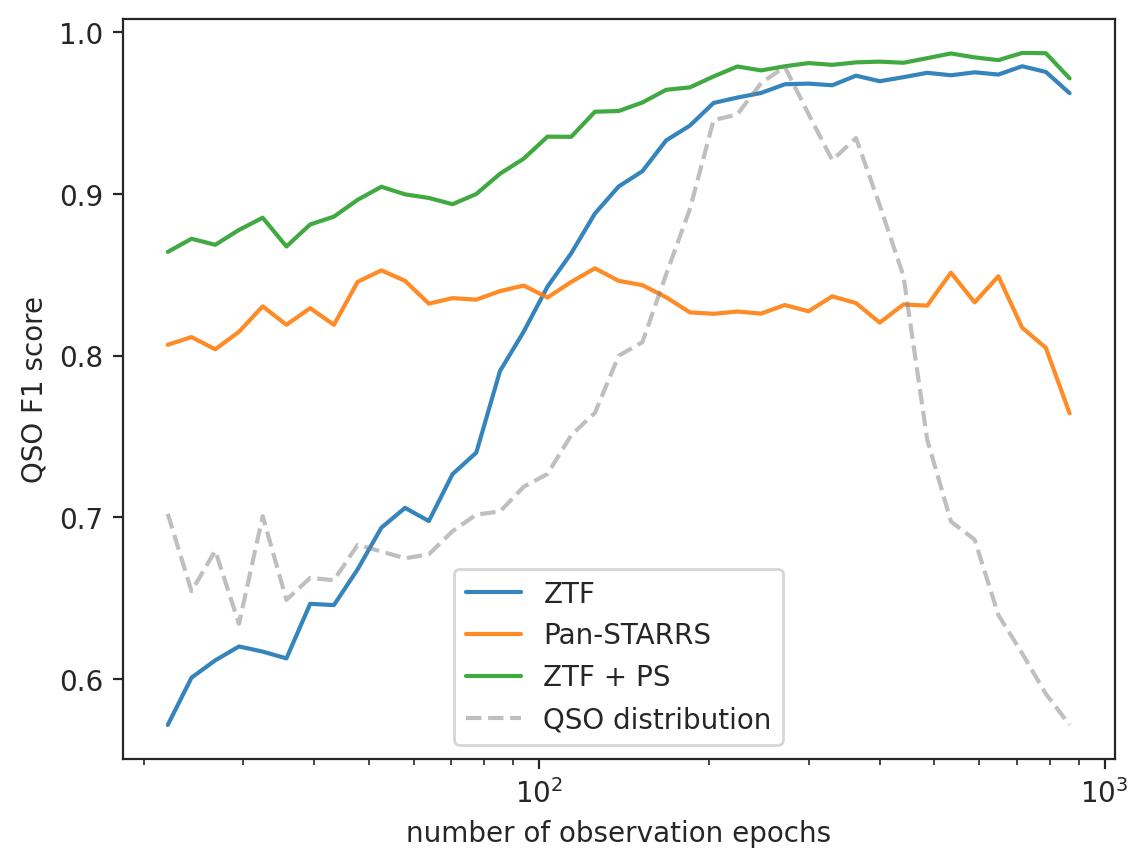}{0.5\textwidth}{}
      \fig{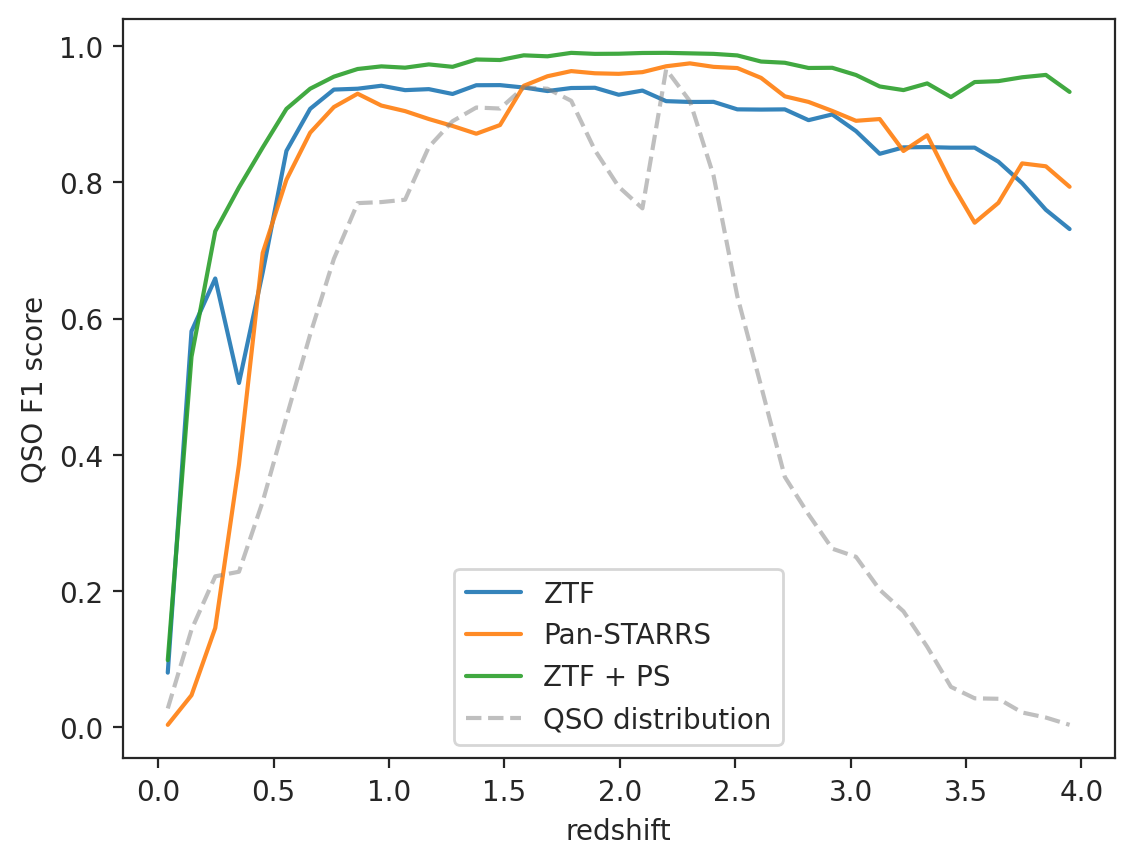}{0.5\textwidth}{}
    }
    \caption{QSO F1 scores for the XGB \textit{g}-band models trained on the ZTF $\times$ PS data, as functions of median magnitude (top left), mean magnitude error (top right), number of ZTF observation epochs (bottom left), SDSS spectroscopic redshift (bottom right). The gray dashed lines show overlaid quasar distributions. The models are trained using combinations of ZTF transformer classification, ZTF magnitude, and PS \textit{griz} magnitudes and colors.}
    \label{fig:results}
\end{figure*}

Since ZTF data span a wide range of magnitudes and observation epochs, we start by analyzing how the classification scores depend on those factors. Fig. \ref{fig:heat} shows the dependence of ZTF \textit{g}-band QSO F1 score on the number of observation epochs and magnitude. We find that given at least 50 observation epochs, the QSO F1 scores deteriorates by 10pp between the objects at magnitude limit $g \sim 20.8$ and the faintest data points at $g \sim 21.5$. We can remove the objects with lower classification quality by applying the dashed red line cut at $g < n_\mathrm{obs} / 80 + 20.375$. Table \ref{tab:cuts} shows the scores for different combinations of limiting magnitude and minimum number of observation epochs. The cuts enable better F1 scores within the catalog, but taking into account the total loss of objects due to cuts, we observe a deterioration in the full sample QSO F1 scores.

\begin{figure}
    \fig{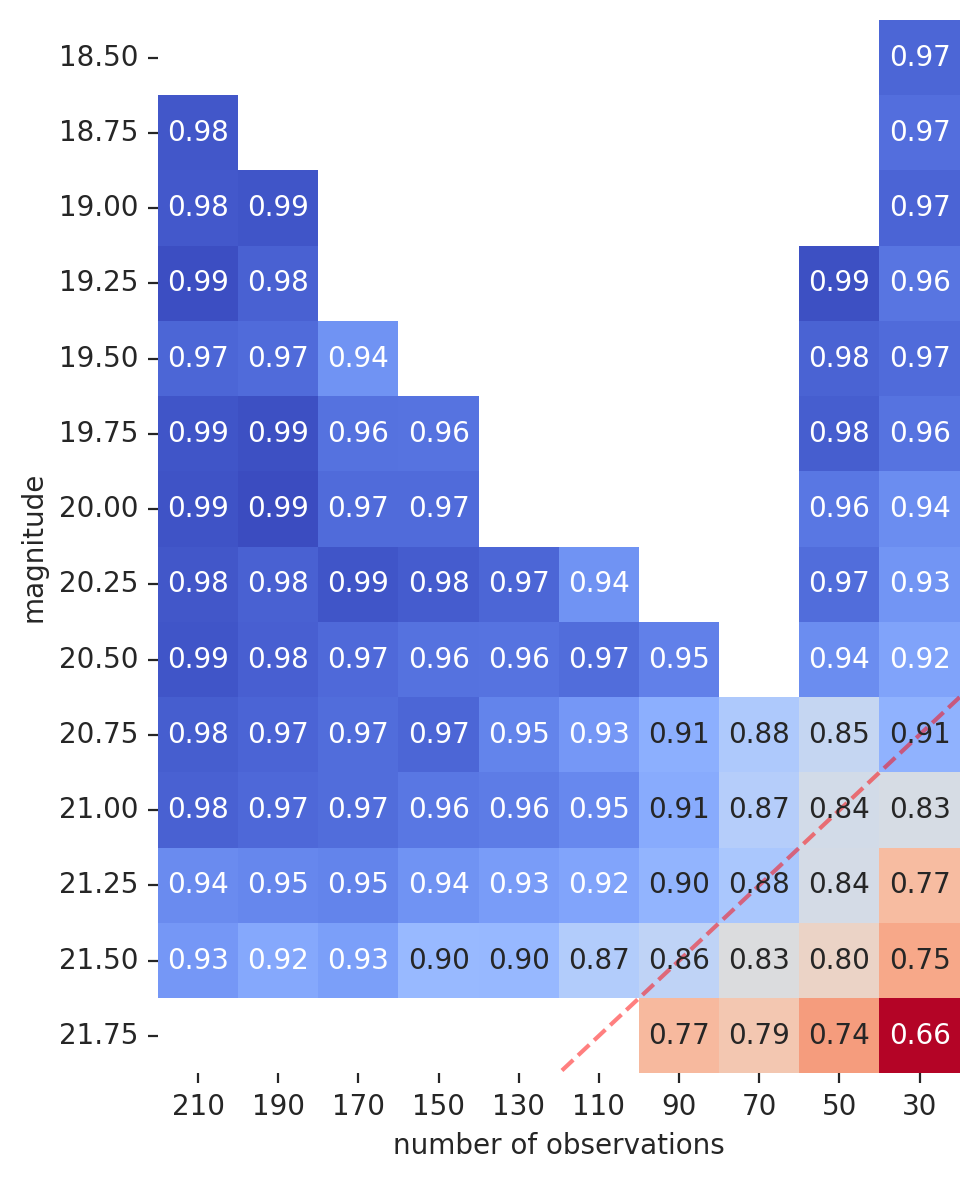}{0.5\textwidth}{}
    \caption{Same as fig. \ref{fig:heat} but for XGB classification using Astromer classification, ZTF \textit{g}-band median magnitudes, and WISE features. This model is used for inference in the final catalog.}
    \label{fig:heat_wise}
\end{figure}

The suggested red line cut, in Figure \ref{fig:heat}, allows to achieve 93\% QSO F1 score, with the full sample metric equal 85\%. The main goal of the cut is to provide classification beyond the $5\sigma$ signal-to-noise limit. We base this cut on the results from ZTF data only, in comparison to adding WISE features as shown later in Fig. \ref{fig:heat_wise}. We want to provide a fiducial version of the catalog with good quality light curves, useful in classification, and with a possible application in search of GW counterparts. It is possible to apply different or more complex cuts, depending on the application. We publish a file with QSOs limited to this cut, and another one with all the ZTF objects classified as quasars.

Fig. \ref{fig:results} shows detailed relations between QSO F1 score, magnitude, noise, number of observation epochs and redshift for the ZTF $\times$ PS training data. We observe that the PS data improves the classification results at the faint end, mostly due to the deeper photometry provided by PS. At brighter magnitudes we observe superiority in ZTF classification over PS. We see that the Astromer's limit of 200 observation epochs does not allow us to probe further increases in scores for more observing epochs. These tests analyze the ZTF classification outputs, but with their weakness that low numbers of objects at certain values, as shown by the gray dashed lines in the Fig. \ref{fig:results}, can result in under-performing models due to a lack of training data with specific characteristics.

\begin{figure*}
    \fig{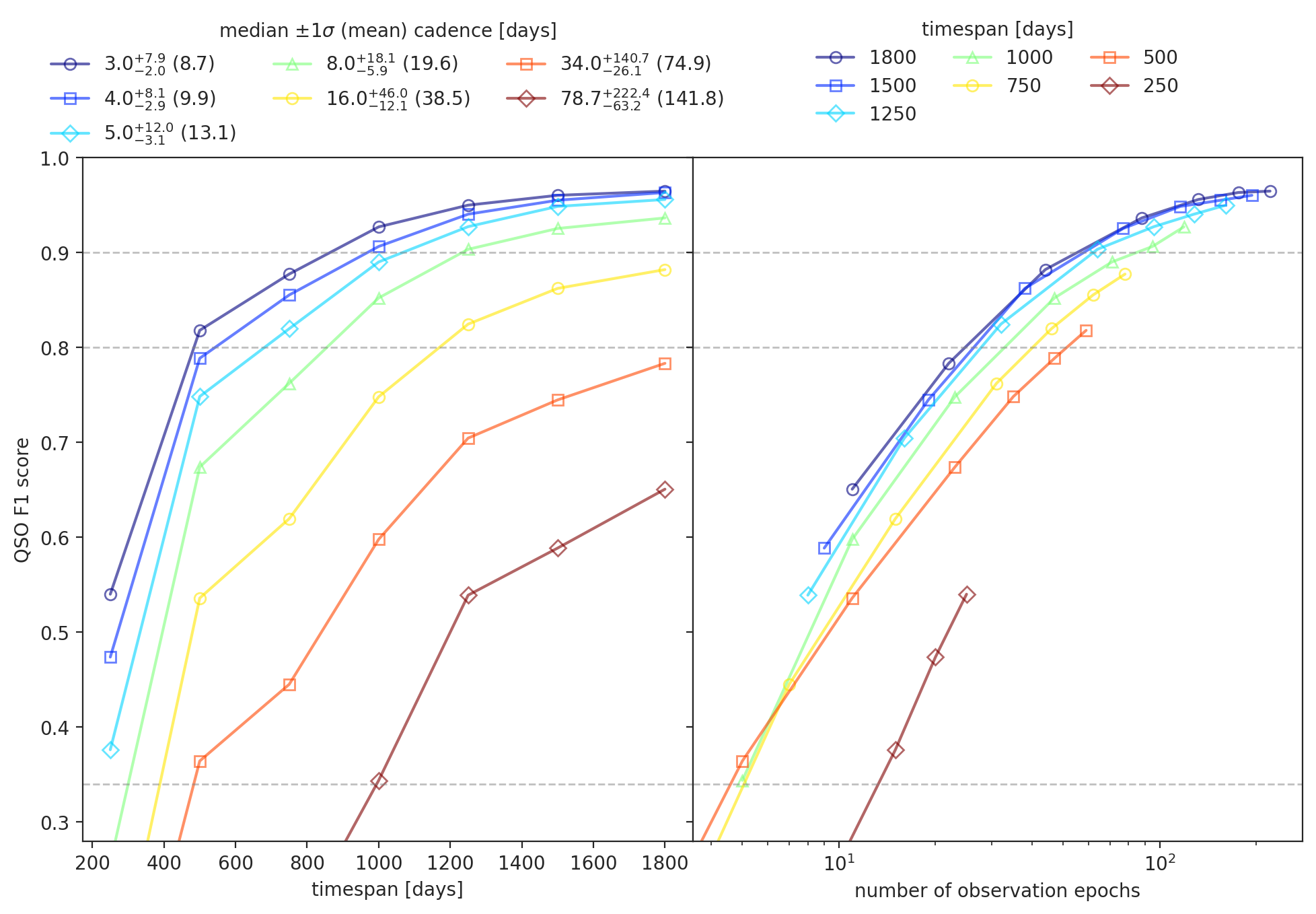}{\textwidth}{}
    \caption{Transformer classification results on the ZTF \textit{g}-band data as a function of limiting magnitude and with subsampled time span and number of observation epochs. Each score corresponds to a new training data and model, each training data has the same size, and each light curve within the given training data has the same time span and number of observations. To obtain lower cadences, we sample 100\%, 80\%, 60\%, 40\%, 20\%, 10\% and 5\% of observation epochs, which corresponds to the successive lines on the left plot. Both plots show the same results: \textit{left}: as a function of time span, and grouped for median cadence, \textit{right}: as a function of number of observations, and grouped for time span.}
    \label{fig:sampling}
\end{figure*}

In order to perform unbiased analysis of relations between classification scores, time span and number of observation epochs, we create training datasets of constant time span and number of observation epochs within all light curves, and train transformer models on each such training set. We create these datasets by taking light curves with time span higher than 1800 days and number of observations higher than the median value. Then, we limit the number of observations to 100\%, 80\%, 60\%, 40\%, 20\%, 10\% and 5\% of the median value, which defines the median cadence. Finally, we limit the timespan by removing specific windows of observation epochs from heads of light curves. Fig. \ref{fig:sampling} shows the results of those tests. We observe strong dependence, and results higher than 90\% obtained either at 900 day time span at the highest 3 day cadence, or the longest 1800 day time span at a lowered 12 day cadence. However, even a 500 day time span or 30 day median cadence can provide results better than 80\%.

We compare the deep learning transformer classification to the shallow ensemble classifier with precomputed features based on ZTF DR5. We observe a difference between 81\% and 89\% QSO F1 scores for the shallow and deep learning models, respectively. However, for a similar setup in ZTF DR20, transformer scores 95\%, as shown in Table \ref{tab:cuts} (row 5), for a magnitude-limited sample. This difference is due to the dataset size and time span difference between DR5 and DR20. The dataset size difference would affect the shallow models less significantly than the deep learning models, hence, we might expect the total difference between shallow and deep learning models to be even larger than the one estimated from DR5.

\subsection{Other surveys} \label{sec:other_surveys}

\begin{deluxetable}{llccc}
\tablenum{2}
\tablecaption{XGB classification results on the ZTF \textit{g}-band data cross-matched with other surveys. We apply limiting magnitude cuts for each survey, and require minimum number of ZTF observation epochs greater than 100. ZTF features include the transformer classifications and \textit{g}-band median magnitude, while Section \ref{sec:other_surveys} describes features from the other surveys.}
\tablewidth{0pt}
\tablehead{
\nocolhead{} & \nocolhead{} & \colhead{QSO} & \colhead{QSO F1} & \colhead{accuracy} \\
\colhead{data} & \colhead{features} & \colhead{fraction} & \colhead{subset} & \colhead{subset}
}
\decimalcolnumbers
\startdata
ZTF           & ZTF                   & 0.5   & 0.97 & 0.96 \\
$\times$ PS         & ZTF                   & 0.49 & 0.97 & 0.96 \\
                  & PS                     &          & 0.82 & 0.90 \\
                  & ZTF $\oplus$ PS         &          & 0.98 & 0.99 \\
$\times$ Gaia    & ZTF                    & 0.48 & 0.97 & 0.97 \\
                  & Gaia                 &          & 0.96 & 0.97 \\
                  & ZTF $\oplus$ Gaia     &          & 0.99 & 0.99 \\
$\times$ WISE   & ZTF                   & 0.36 & 0.97 & 0.97 \\
                 & WISE                 &         & 0.98 & 0.93 \\
                 & ZTF $\oplus$ WISE    &         & 0.99 & 0.99 \\
\enddata
\tablecomments{\quotes{QSO fraction} provides the fraction of QSOs present in the cross-matches with respect to the full ZTF training sample, and \quotes{accuracy} stands for the general three class accuracy.}
\label{tab:surveys}
\end{deluxetable}

We use PS, WISE and Gaia to compare the light curve classification with other types of surveys. Table \ref{tab:surveys} presents a detailed comparison between the ZTF, PS, WISE and Gaia surveys. The results are calculated on the test data with applied limiting magnitudes and require at least 100 ZTF observation epochs in order to minimize the magnitude differences between the surveys and compare their full predictive power. The comparison between ZTF and PS surveys shows 97\% and 82\% QSO F1 score, in favor of the light curve based classification. The results between ZTF, Gaia and WISE are similar, however, each of those surveys gains significant improvement when merging with the ZTF observations.

\begin{deluxetable}{lcc}
\tablenum{3}
\tablecaption{XGB classification results on the ZTF \textit{g}-band training data, with number of observation epochs greater than 20. In contrast to Table \ref{tab:surveys}, here the XGB models classify objects with missing features from the other surveys.}
\label{tab:final}
\tablewidth{0pt}
\tablehead{
\colhead{features} & \colhead{QSO F1} & \colhead{accuracy}
}
\decimalcolnumbers
\startdata
ZTF                                          & 0.88 & 0.88 \\
ZTF $\oplus$ PS                                & 0.95 & 0.96 \\
ZTF $\oplus$ WISE                          & 0.94 & 0.94 \\
ZTF $\oplus$ Gaia                            & 0.91 & 0.95 \\
ZTF $\oplus$ PS $\oplus$ WISE                & 0.96 & 0.97 \\
ZTF $\oplus$ PS $\oplus$ Gaia                  & 0.96 & \textbf{0.98} \\
ZTF $\oplus$ PS $\oplus$ WISE $\oplus$ Gaia  & \textbf{0.97} & \textbf{0.98} \\
PS $\oplus$ WISE $\oplus$ Gaia              & 0.95 & 0.97 \\
\enddata
\tablecomments{\quotes{Accuracy} stand for the general three class accuracy. Boldface shows the highest value in each column.}
\end{deluxetable}

Table \ref{tab:final} shows the classification results of the full ZTF training sample, with features from other surveys added where available, and XGB processing the missing features where present. Adding PS, WISE, and Gaia features to the ZTF classification improves the QSO F1 scores by 7, 6, and 3 percent points, respectively. As shown in Section \ref{sec:lc_classification}, in case of PS, the improvement is mostly at the faint end due to the PS fainter limiting magnitude. The last row shows that adding ZTF data to all the other surveys increases the classification score by 2 percent points.

\begin{figure}
    \gridline{
        \fig{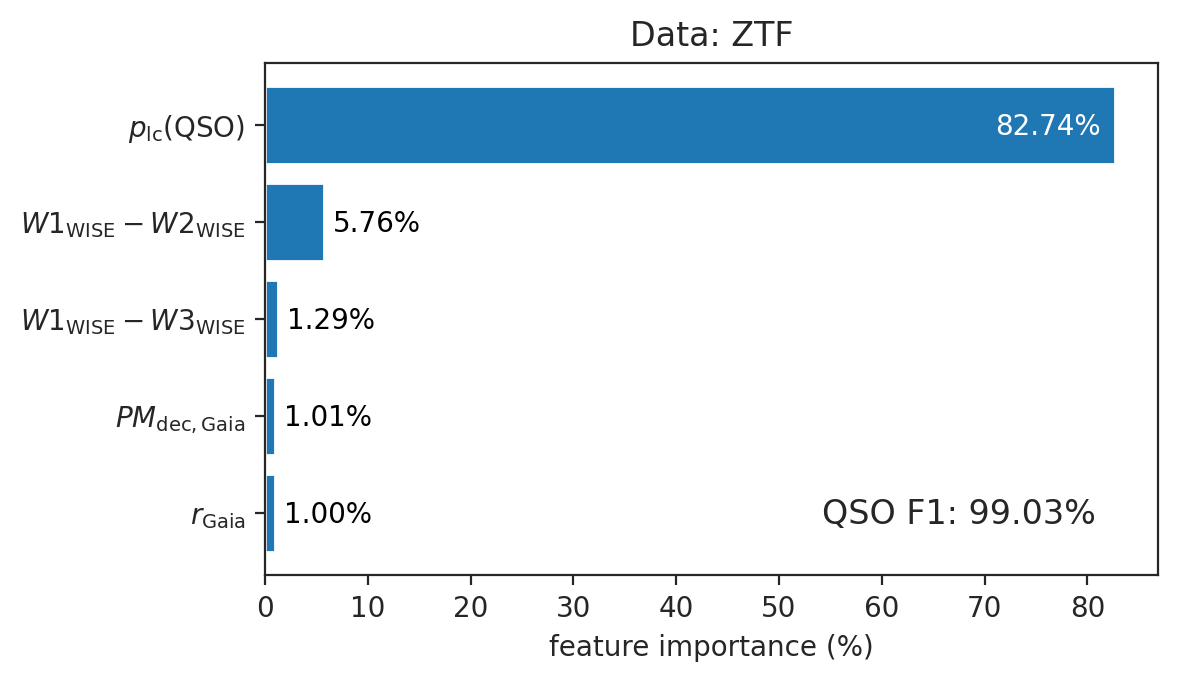}{0.5\textwidth}{}
    }
    \gridline{
        \fig{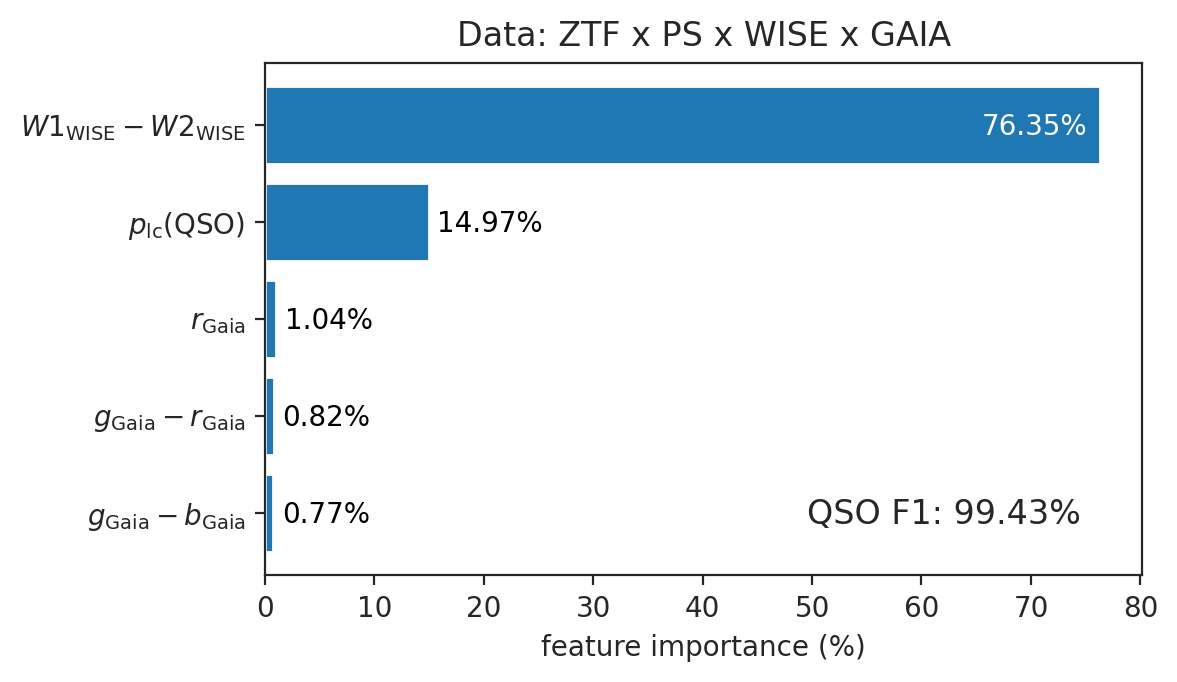}{0.5\textwidth}{}
    }
    \caption{Feature importance from XGB models trained on binary QSO vs non-QSO classification problem, using the ZTF \textit{g}-band data with included features from the other surveys. \textit{Top}: ZTF training sample and processing of missing features, \textit{bottom}: subsample of objects detected by every survey, with all the features present. $p_{\mathrm{lc}}\mathrm{(QSO)}$ stands for the ZTF light curves based QSO classification probability.}
    \label{fig:features}
\end{figure}

Fig. \ref{fig:features} presents feature rankings from XGB models trained on all available features in two cases: all the ZTF data with processing of missing features, and a subset of objects detected by all surveys, with no missing features present. The rankings are based on binary QSO vs non-QSO classification to focus on the features important for QSO detection. The full ZTF sample measures the importance with regard to all available QSOs, but may potentially underestimate WISE importance, which is available for 72\% of QSOs. On the other hand, the cross-match between all the surveys measures the feature importance only for red and point-like objects detectable by both the WISE and Gaia surveys. The top plot shows that the transformer's QSO probability is the most important feature. The bottom plot shows that for the cross-matched data, $W1-W2$ color is the most important feature, but the transformer classification takes second place with 15\% importance. Within the given set of surveys, parallax and proper motion measurements from Gaia are not the most important features.

\subsection{Redshift} \label{sec:redshift}

\begin{figure}
    \fig{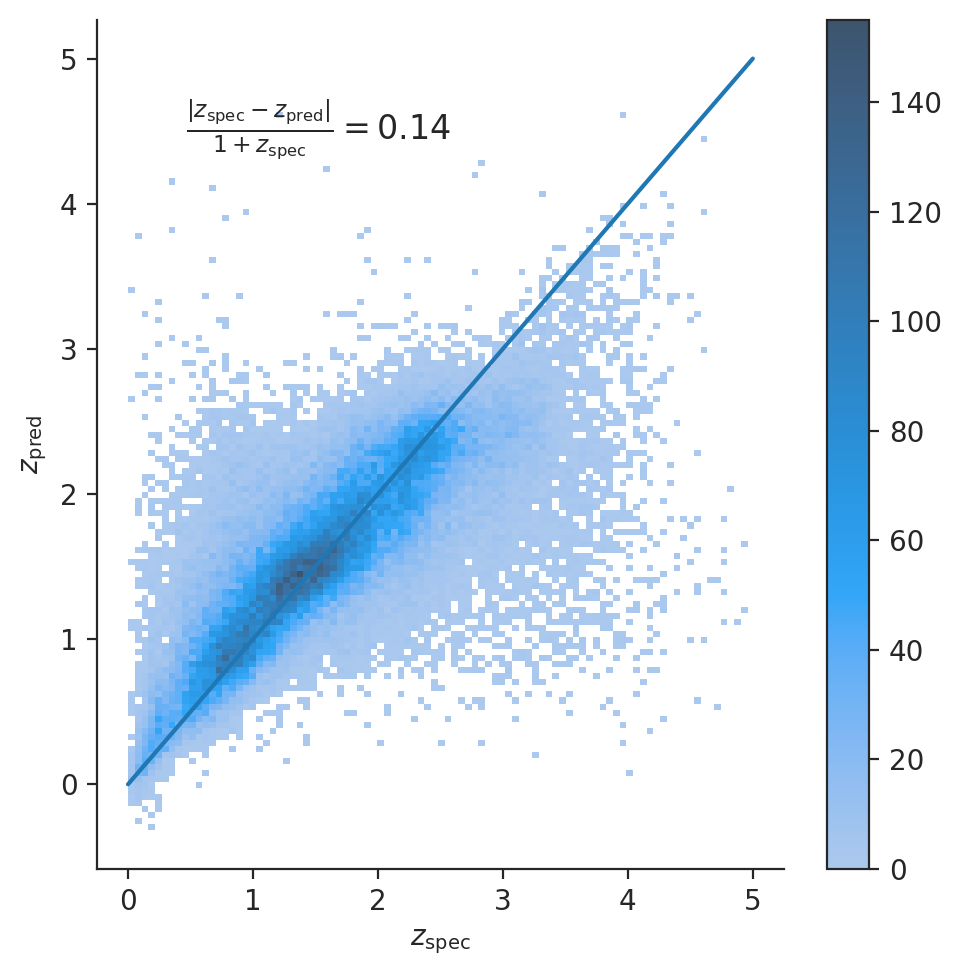}{0.5\textwidth}{}
    \caption{Predicted redshift estimates, $z_{\mathrm{pred}}$, for the XGB model trained on ZTF $\times$ WISE data, vs spectroscopic redshift, $z_{\mathrm{spec}}$, limited to QSOs with at least 20 observation epochs. The model uses WISE magnitudes and colors, as well as ZTF \textit{g}-band magnitude.}
    \label{fig:redshift}
\end{figure}

We find that the light curves do not allow us to estimate redshifts for the quasars, and within the analyzed surveys, the WISE data are necessary to obtain the redshift estimates. Hence, for objects with available WISE data, we publish redshift estimates using XGB model trained on ZTF $\times$ WISE data. Our measured error is $\Delta z/(1 + z) = 0.14$, with percentage of predictions more than 20\% and 50\% different then the spectroscopic redshift value equal 42\% and 14\%, respectively. The model uses ZTF \textit{g}-band magnitude and WISE colors, but WISE colors are the most important features. Figure \ref{fig:redshift} shows the results of redshift estimation on the ZTF $\times$ WISE test data.

\begin{figure*}
    \gridline{
      \fig{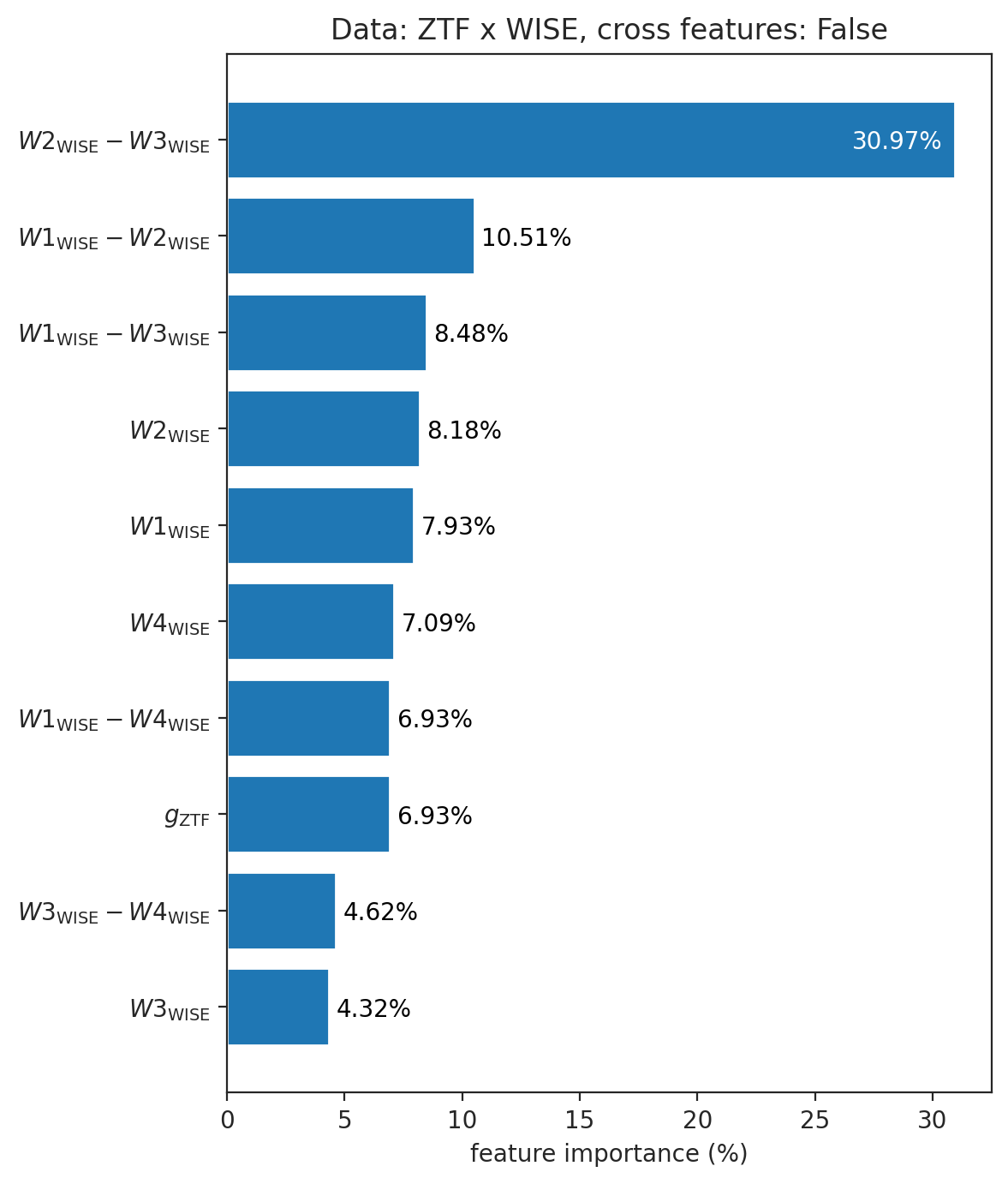}{0.5\textwidth}{}
      \fig{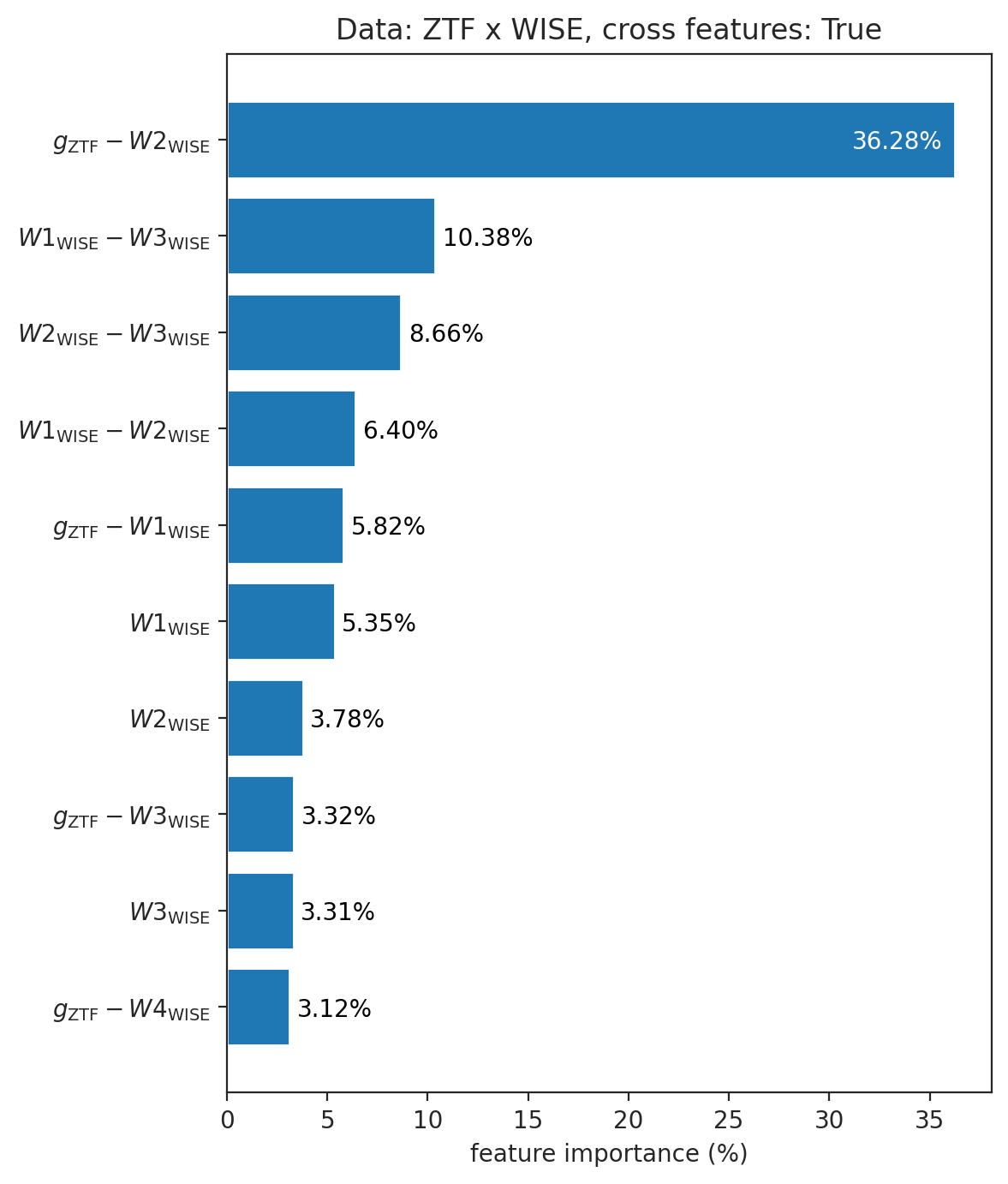}{0.5\textwidth}{}
    }
    \caption{Feature importance for QSO photometric redshifts from XGB models trained on the ZTF \textit{g}-band and WISE data. \textit{Left}: colors from standalone surveys only, \textit{right}: added features from a mix of ZTF and WISE surveys. Both approaches provide the same redshift estimation error.}
    \label{fig:features_redshift}
\end{figure*}

Fig. \ref{fig:features_redshift} shows that all WISE bands are useful for QSO redshifts. We consider adding colors from a mix of ZTF and WISE survey, such as $g_{ZTF} - W1_{WISE}$. However, both approaches provide similar performance. Adding features which do not improve the estimates may lead to overfitting, hence, we decide not to add such colors in the final inference model. We check that requiring W[1-2] bands provides 323M object in the ZTF inference dataset, while requiring W[1-4] bands provides 262k less objects, which is 0.08\% difference. Due to that, we use all WISE bands in the inference.

\subsection{The QZO catalog} \label{sec:catalog}

\begin{figure}
    \fig{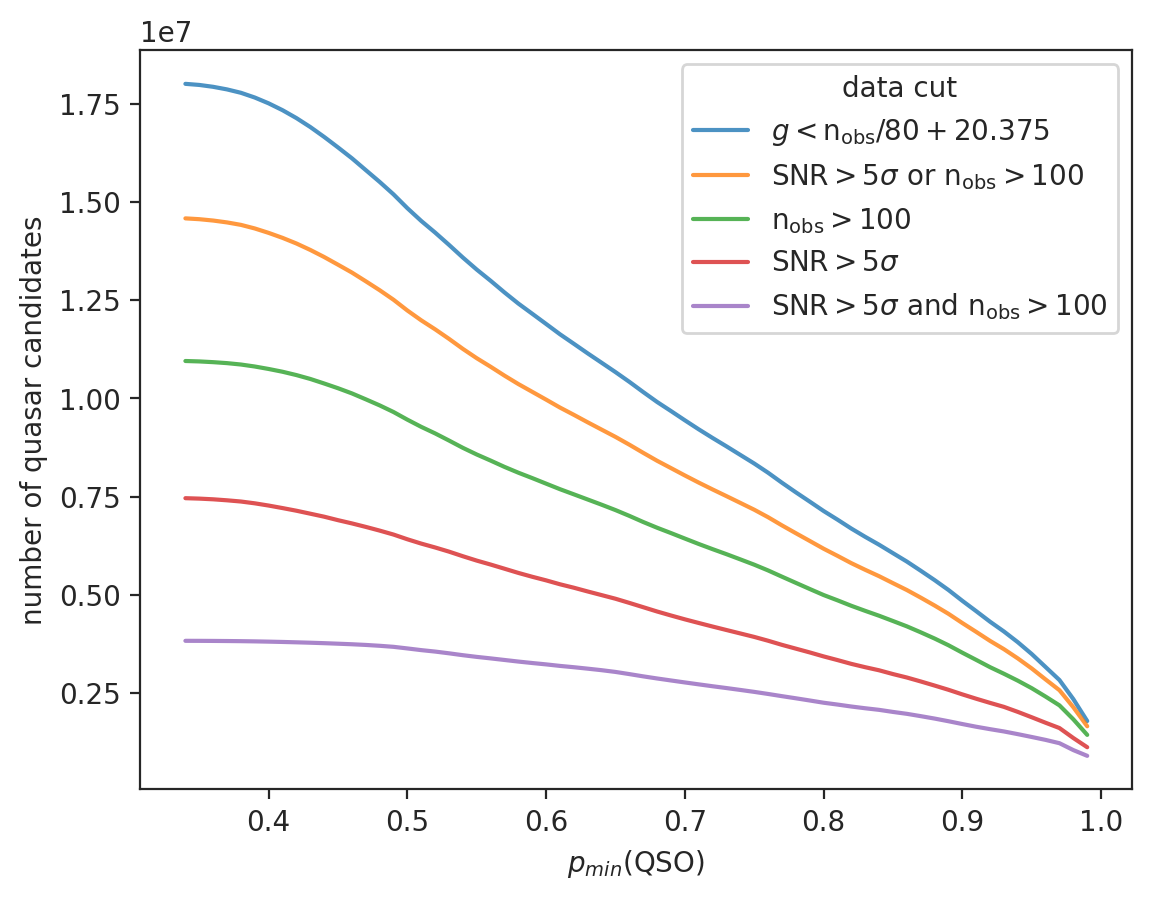}{0.5\textwidth}{}
    \caption{Number of QSO candidates from inference on the ZTF \textit{g}-band data with XGB trained on ZTF and WISE features, as a function of minimum QSO classification probability, and grouped for different data cuts.}
    \label{fig:counts}
\end{figure}

\begin{deluxetable*}{lcccccccc}
\tablenum{4}
\tablecaption{Number of quasar candidates, in millions, from inference on the ZTF \textit{g}-band data with XGB models trained using ZTF and WISE features, at different cuts on data and minimum QSO classification probability.}
\label{tab:counts}
\tablewidth{0pt}
\tablehead{
\nocolhead{} & \nocolhead{} & \nocolhead{} & \multicolumn6c{number of QSOs at $p_{\mathrm{QSO}}$ higher than} \\
\colhead{data cut} & \colhead{QSO F1} & \colhead{accuracy} & \colhead{0.3} & \colhead{0.5} & \colhead{0.8} & \colhead{0.9} & \colhead{0.95} & \colhead{0.99}
}
\decimalcolnumbers
\startdata
    none                                            & 0.94 & 0.94 & 45.93 & 32.23 & 11.26 & 6.64 & 4.50 & 2.02 \\
    $g < n_\mathrm{obs} / 80 + 20.375$           & 0.97 & 0.96 & 18.01 & 14.84 & 7.14 & \textbf{4.85} & 3.50 & 1.79 \\
    $\mathrm{SNR} > 5\sigma \lor n_\mathrm{obs} > 100$      & 0.97 & 0.96 & 14.59 & 12.24 & 6.18 & 4.28 & 3.13 & 1.66 \\
    $n_\mathrm{obs} > 100$                                & 0.97 & 0.96 & 10.96 & 9.46 & 5.00 & 3.53 & 2.63 & 1.44 \\
    $\mathrm{SNR} > 5\sigma$                                & 0.98 & 0.97 & 7.46 & 6.42 & 3.43 & 2.47 & 1.89 & 1.12 \\
    $\mathrm{SNR} > 5\sigma \land n_\mathrm{obs} > 100$      & 0.98 & 0.98 & 3.83 & 3.64 & 2.26 & 1.71 & 1.38 & 0.90 \\
\enddata
\tablecomments{\quotes{Accuracy} stands for the general three class accuracy, and both \quotes{QSO F1} and \quotes{accuracy} describe the test data results for a given setup of data cuts. SNR cut is applied to the \textit{g}-band only. Boldface shows the fiducial setup.}
\end{deluxetable*}

Based on the results from Sections \ref{sec:other_surveys} and \ref{sec:redshift}, and taking into account computational complexity of including more surveys in the final inference, we publish the classification results based on the ZTF and WISE surveys, together with the ZTF $\times$ WISE redshift estimates. We also publish the XGB and ANN models, which can be used to create a catalog based on a different combination of surveys.

In the full ZTF catalog, we remove light curves which within $1\arcsec$ have at least one neighbor with more observation epochs. This step removes duplicated light curves by choosing the ones with the highest number of observation epochs. Figure \ref{fig:counts} and Table \ref{tab:counts} show the resulting number of objects classified as QSOs at different data and classification probability cuts. The full catalog consists of 45.93 million objects classified as QSOs, but due to the class and magnitude distribution differences between the training and inference data, it is necessary to calibrate the resulting classification probabilities. If such calibration is not possible, as in our case, one can establish a minimum classification probability to reduce QSO contamination by stars and galaxies, at a cost of lowered QSO completeness. From Section \ref{sec:lc_classification} we conclude that a high quality QSO classification requires a cut based on magnitude and number of observation epochs at $g < n_\mathrm{obs} / 80 + 20.375$, which we consider a fiducial cut. At this cut, the estimated QSO F1 score equals 97\%. \cite{Nakoneczny_2019, Nakoneczny_2021} uses a test based on Gaia parallaxes for a similar task of QSO classification based on SDSS data and suggests 90\% minimum QSO classification probability. We consider this as an additional fiducial cut, while other cuts are possible depending on the application. We publish all the objects classified as quasars to allow for any postprocessing of the predictions. We obtain a limited catalog of higher purity by applying both the data and classification probability cuts, resulting in 4,849,574 objects classified as QSOs. We call this catalog QZO. We also expect many correctly classified QSOs in the fainter sample below red line in fig \ref{fig:heat}, but with an unknown and higher contamination rate by stars and galaxies, as shown in Fig. \ref{fig:heat} and \ref{fig:heat_wise}. The photometric redshifts are available for 33\% of the QZO objects with matching W[1-4] bands from AllWISE.

WISE coverage equals 41\% and 77\% for the inference and training data, respectively. The difference might constitute a problem, if the missing features are due to a missing sky coverage in inference data, while ML model learns on the ZTF and SDSS cross-matched footprint that missing WISE features are due to their physics and not being visible in the WISE infrared observations. We consider predictions from two XGB models, one taught on light curve classification and ZTF median magnitude, and a second one with WISE features added. The second model is the one which we use to create the QZO catalog. Among the objects with no WISE observations, and using the same cuts as in the case of QZO catalog, there are 2.8M objects classified as quasars by both of these models. However, there are also 0.65M objects classified as QSOs only by the model which does not use WISE features, and 0.31M objects classified as such only by the model which includes the WISE features, despite the fact that WISE data are only missing entries in this comparison. It is not obvious which one of these smaller samples should be accepted. The model without WISE features performs worse generally, but the model with WISE features can fail due to a misinterpretation of a meaning of missing WISE features. In the published files, described in Appendix \ref{app:catalog}, we published the 0.31M sample as part of the QZO catalog, and the 0.65M sample in the file containing all objects classified as QSOs by any of the two ML models.

The difference between training and inference data in WISE coverage, as well as in distribution of magnitude and number of observations (Fig. \ref{fig:data}), can be taken into account to calibrate the QSO F1 score of the final catalog. The score can be calculated on the training data as a function of variables of interest, and then calibrated using an inference distribution of these variables. It can be achieved by cross-matching our full catalog back with SDSS, analyzing the quality of predictions on the training subset, and then calibrating the scores for any given subset of the catalog. We focus on describing the differences between the training and inference data, and how these affect the inference quality, while leaving the calibration efforts to any particular use case of our catalog.

\begin{figure}
    \fig{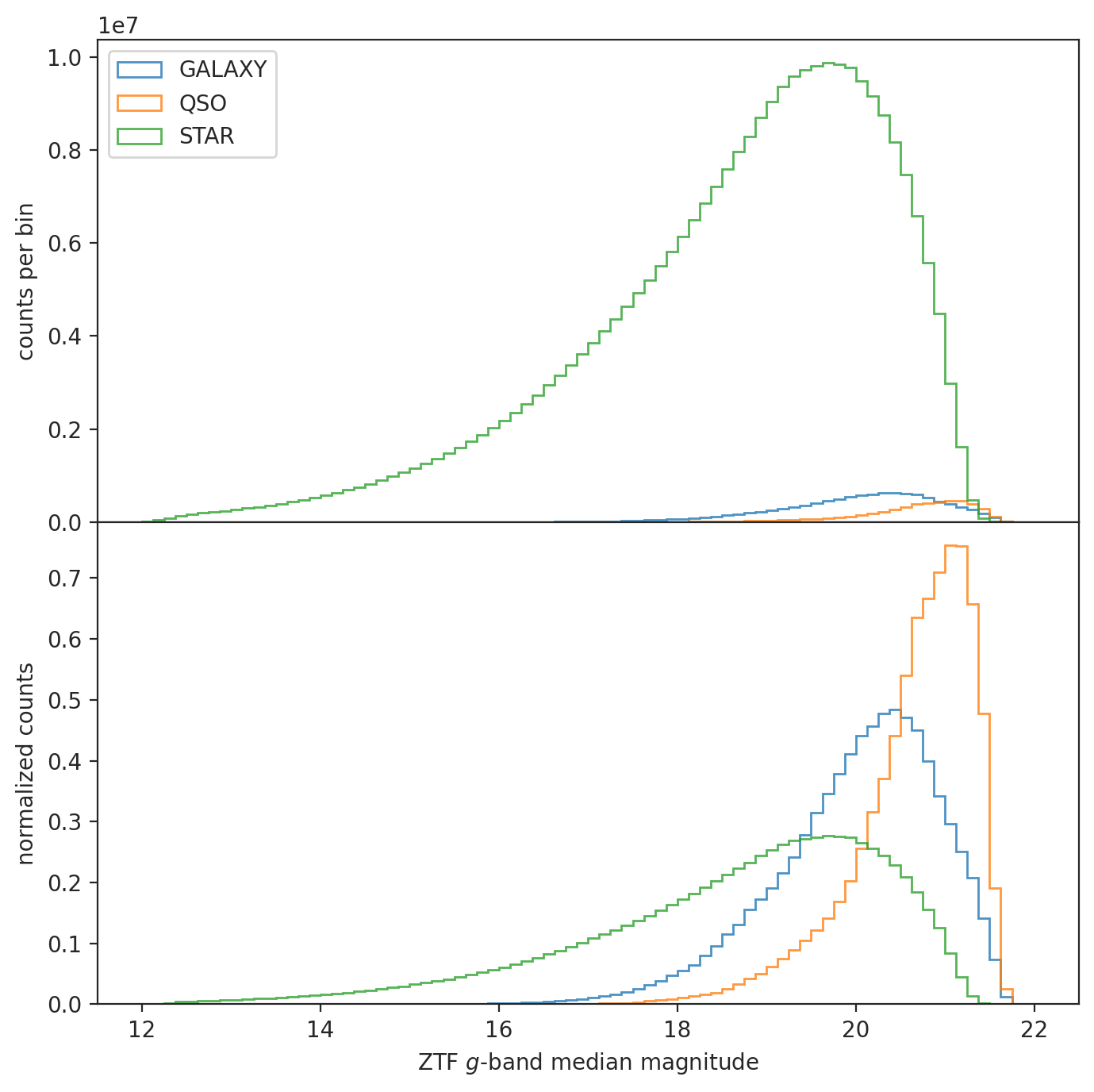}{0.5\textwidth}{}
    \caption{Magnitude distribution based on the inference with XGB model including WISE features, with each class limited to classification probability higher than 90\%, and applying the fiducial magnitude and number of observation cut. \textit{Top}: counts per bin, \textit{bottom}: areas normalized to unity.}
    \label{fig:mag_inference}
\end{figure}

Fig. \ref{fig:mag_inference} shows magnitude distributions for each class in the inference data, based on the XGB model with WISE features, limited to classification probability higher than 90\%, and applying the fiducial magnitude and number of observation cut. We observe the model mostly classifying objects as stars, and number of quasars rising with increasing magnitude, as noticed earlier in Fig. \ref{fig:data}. Additionally, we see that the bimodal galaxy distribution present in the training data does not bias the final inference.

\begin{figure}
    \fig{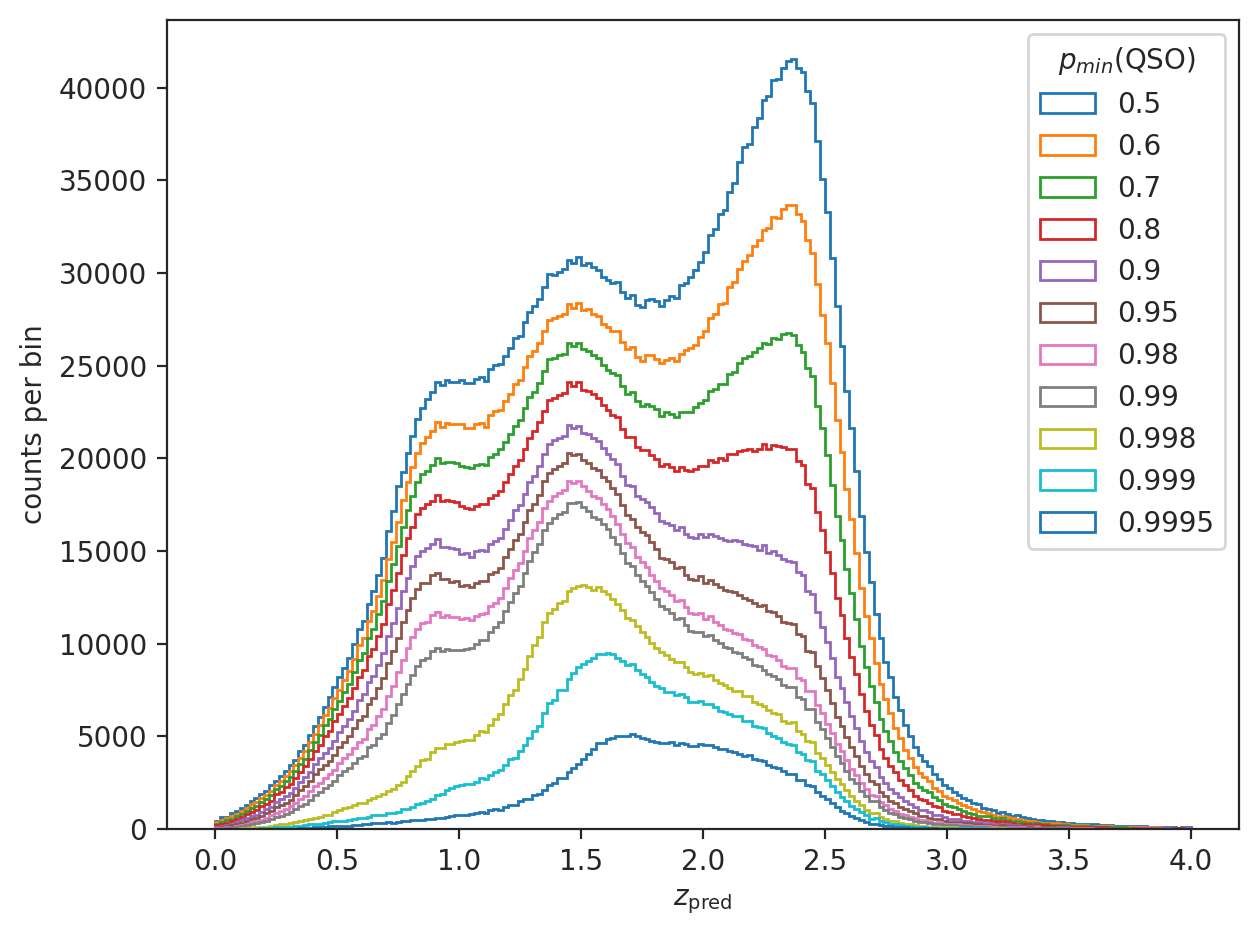}{0.5\textwidth}{}
    \caption{Photometric redshift distribution for a subsample of the quasar catalog with WISE features available, at $g < n_\mathrm{obs} / 80 + 20.375$, as a function of minimal QSO classification probability.}
    \label{fig:catalog_redshift}
\end{figure}

Fig. \ref{fig:catalog_redshift} shows photo-z distribution for quasars with available WISE features at fiducial $g < n_\mathrm{obs} / 80 + 20.375$. At lowest quasar classification probability, we observe a maximum of photo-z estimates at $z \sim 2.4$. The fiducial cut at $p_\mathrm{QSO} > 0.9$ shows a maximum at $z \sim 1.5$, with broad wings of the distribution reaching to $z \sim 0.8$ and $z \sim 2.4$. An influx of QSOs at $z \sim 0.8$ disappears at $p_\mathrm{QSO} > 0.998$.

\section{Palomar spectroscopy} \label{sec:observations}

\begin{deluxetable*}{lcccccccc}
\tablenum{5}
\tablecaption{List of random three SDSS galaxies and stars at ZTF \textit{g}-band magnitude limit and number of observation epochs higher than 100, classified as AGN by our deep learning model. The $z_\mathrm{pred}$ column is additionally based on WISE features.}
\tablewidth{0pt}
\tablehead{
\multicolumn3c{SDSS} & \multicolumn5c{ZTF} & \colhead{notes} \\
\colhead{name} & \colhead{z} & \colhead{class} & \colhead{AGN} & \colhead{galaxy} & \colhead{star} & \colhead{light curve} & \colhead{$z_\mathrm{pred}$} & \nocolhead{}
}
\decimalcolnumbers
\startdata
J224649.18+132545.9 & 0.23 & galaxy & 45.8\% & 39.1\% & 15.1\% & fig. \ref{fig:galaxy_as_qso} \textit{b} & 0.35 & AGN; early-type galaxy with weak [\ion{N}{2}] \\
J230105.96+275256.4 & 0.28 & galaxy & 48.3\% & 35.5\% & 16.2\% & fig. \ref{fig:galaxy_as_qso} \textit{e} & 0.49 & AGN; early-type galaxy with weak [\ion{N}{2}] \\
J004038.40+160949.9 & 0.29 & galaxy & 99.7\% & 0.3\% & 0.1\% & fig. \ref{fig:galaxy_as_qso} \textit{g} & 0.45 & AGN; broad H$\alpha$ emission \\
J002411.35-003310.9 & - & star & 60.5\% & 3.5\% & 36.1\% & fig. \ref{fig:star_as_qso} \textit{b} & - & K5 star; MgIb and NaD absorption \\
J155654.47+210719.0 & - & star & 71.5\% & 9.7\% & 18.9\% & fig. \ref{fig:star_as_qso} \textit{e} & - & CV; with strong Balmer emission \\
J185111.24+174645.7 & - & star & 74.1\% & 2.5\% & 23.4\% & fig. \ref{fig:star_as_qso} \textit{g} & - & F9 star; MgIb and NaD absorption \\
\enddata
\tablecomments{First two objects were observed at Palomar on 2024 Sep 11 by Daniel Stern, with 2x1200s exposure time, and clear 1.3 arcsec seeing.}
\label{tab:palomar}
\end{deluxetable*}

We randomly selected six sources with conflicting SDSS and ZTF light curve classifications as a pilot program at Palomar Observatory on a night assigned for other science.  The sample included three sources identified by SDSS as galaxies based on their spectra and three sources identified by SDSS as Galactic. All six were classified as AGN by our model. Table \ref{tab:palomar} presents these six sources, and they can be also identified in figures \ref{fig:galaxy_as_qso} and \ref{fig:star_as_qso}. All three SDSS Galactic sources were clearly Galactic based on the SDSS spectra, while one of the SDSS inactive galaxies was clearly active based on the SDSS spectrum.

We obtained optical spectroscopic follow-up of the other two extragalactic sources identified as inactive by SDSS, but active from the machine learning analysis of their ZTF light curves.  The observations were obtained using the Double Spectrograph \citep[DBSP, ][]{Oke_1982} on the 5m Hale telescope at Palomar Observatory on UT 2024 September 11, which was a photometric night with $\sim 1\farcs3$ seeing.  For both sources, we obtained two exposures of 1200~s using the $1\farcs5$ slit, the 600 line blue grating (blazed at 4000~\AA), the 5500~\AA\, dichroic, and the 316 line red grating (blazed at 7500~\AA).  The slits were aligned at the parallactic angle and the data were reduced using standard techniques within IRAF.

The Palomar spectra of both sources are dominated by stellar populations, showing early-type galaxies with strong absorptions from calcium H and K, the G-band, MgIb, and NaD, as well as prominent 4000~\AA\, breaks. Both galaxies also show weak [\ion{N}{2}]~6584 emission, indicative of a buried AGN.  The SDSS spectra also show weak evidence for this nitrogen emission, which is confirmed by the Palomar data.  No other strong emission lines are evident. As shown in Table~\ref{tab:palomar}, our model provides a probabilistic likelihood for a source being active, rather than the binary SDSS classifications. The two sources were predicted to be active at the 46-48\%\, likelihood, and inactive at the 36-39\%\, likelihood.  Their ZTF light curves show variability by a few tenths of a magnitude since 2019 (fig. \ref{fig:galaxy_as_qso}, panels \textit{b} and \textit{e}), which is somewhat surprising since the spectroscopy implies they are heavily buried, or type-2, AGN.  Their optical light is expected to be dominated by their stellar populations.  Indeed, the two galaxies have WISE W1-W2 colors of 0.25-0.28, indicating that their mid-IR emission are also dominated by their stellar populations \citep[e.g., ][]{Stern_2012}.

The third SDSS galaxy candidate, J004038.40+160949.9, was classified with a 99.7\%\ likelihood of being active and has varied by $\sim 0.8$~mag since 2019.  The SDSS spectrum, though classified as a starburst galaxy, shows clear broad H$\alpha$ emission and thus was not re-observed with Palomar.  The first SDSS Galactic candidate, J155654.47+210719.0, shows minimal variability other than a strong flare in early 2020.  The SDSS spectrum shows strong hydrogen Balmer emission and is correctly classified as a cataclysmic variable.  The ML classifier identified this source with an 18.9\% likelihood of being Galactic and a 71.5\% likelihood of being a quasar.  The other two stars show minimal variability, $\lesssim 0.2$~mag, and are clearly normal Galactic stars based on their SDSS spectroscopy.

In summary, from this small sample, all three sources identified as Galactic by SDSS but as quasars by the ML algorithm were indeed Galactic, while all three sources identified as inactive galaxies by SDSS  but as quasars by the machine learning algorithm were indeed active.

\section{Conclusion} \label{sec:conclusion}

We publish a catalog of reliable QSO candidates based on time domain observations, created using deep learning classification of ZTF light curves with a transformer model, and then ensembled by XGB with WISE survey data. The high quality sample consists of 4,849,574 QSOs, with an estimated QSO F1 score of 97\%. For 33\% of QZO objects, with available WISE data, we publish redshifts with estimated error $\Delta z/(1 + z) = 0.14$. We show the superiority of light curve based classification over static \textit{griz} magnitudes and colors, and we reach similar classification capabilities to the WISE and Gaia surveys. However, both WISE and Gaia gain a significant improvement when optical light curves are added into their classifications. We also show that the ZTF light curve classifications are the most important features for QSO detection. In the case of red point-like objects detectable by both the Gaia and WISE surveys, $W1-W2$ is the most important feature, but the addition of light curve classification also improves the classification. We find that given at least 50 observation epochs, the QSO F1 scores deteriorate by 10pp between the objects at $5\sigma$ SNR magnitude limit $g \sim 20.8$ and the faintest data points at $g \sim 21.5$. We robustly classify objects fainter than the magnitude limit by requiring $g < n_\mathrm{obs} / 80 + 20.375$. At the ZTF 3 day median cadence, a $\geq 900$ days time span is required to obtain a QSO F1 score higher than 90\%. However, at the highest available time span of 1800 days, even a 12 day median cadence can provide similar results. Additionally, even a 500 day time span or 30 day median cadence can provide results higher than 80\%.

We perform the analysis with the pre-trained Astromer transformer model. We show that the deep learning methods are necessary to fully optimize the classification with time-domain data. We successfully re-train the transformer model from ZTF DR10, on which it was originally trained, to DR20, and also successfully perform transfer learning from the ZTF \textit{g}- to \textit{r}-band. The usage of a pre-trained model significantly lowers the computational time, and our approach of re-training the encoder twice, first in the semi-supervised way on longer light curves, and then in the supervised manner to the SDSS classification, further improves the cores and the final training time, especially considering a vast number of re-training experiments performed. The Astromer is limited to a maximum of 200 observation epochs per light curve, and we show that the scores might improve further with longer inputs. Therefore, we conclude that our work might still not fully exploit the ZTF capabilities.

The analysis of a small sample of incorrectly classified light curves shows that the ZTF classification can even correct misclassified SDSS objects. Considering this, and the high scores obtained by our models, the next steps of the analysis involve a more nuanced approach to the QSO light curves. The possibilities include a supervised and unsupervised classification of QSOs and their subtypes, as well as a fully unsupervised search for anomalies with deep learning, using the 4.8M of objects from the QZO catalog.

A very important aspect of our analysis is that it was performed on real time domain data, which are irregularly sampled and span a wide range of available observation epochs. Our results show how much information can be extracted from the time domain surveys with deep learning. We make important conclusions from processing of the time domain data with deep learning, which can be used to help design future large time domain surveys.  The catalog is ready for numerous QSO applications, and provides an important step towards a systematic search of electromagnetic counterparts to the gravitational wave events embedded in QSO accretion disks.

\section*{Data availability}

We publish the catalog and models on Zenodo \citep{QZO_data_doi}, while the code is available on Zenodo \citep{QZO_soft_doi}, and GitHub \footnote{\url{https://github.com/snakoneczny/ztf-agn}}.

\begin{acknowledgments}

SJN and MJG are supported by the US National Science Foundation (NSF) through grant AST-2108402.
Based on observations obtained with the Samuel Oschin Telescope 48-inch and the 60-inch Telescope at the Palomar Observatory as part of the Zwicky Transient Facility project. ZTF is supported by the National Science Foundation under Grants No. AST-1440341 and AST-2034437 and a collaboration including current partners Caltech, IPAC, the Oskar Klein Center at Stockholm University, the University of Maryland, University of California, Berkeley , the University of Wisconsin at Milwaukee, University of Warwick, Ruhr University Bochum, Cornell University, Northwestern University and Drexel University. Operations are conducted by COO, IPAC, and UW.
Wide-field Infrared Survey Explorer, which is a joint project of the University of California, Los Angeles, and the Jet Propulsion Laboratory/California Institute of Technology, and NEOWISE, which is a project of the Jet Propulsion Laboratory/California Institute of Technology. WISE and NEOWISE are funded by the National Aeronautics and Space Administration.
European Space Agency (ESA) mission {\it Gaia} (\url{cosmos.esa.int/gaia}), processed by the Gaia Data Processing and Analysis Consortium (DPAC, \url{cosmos.esa.int/web/gaia/dpac/consortium}). Funding for the DPAC has been provided by national institutions, in particular the institutions participating in the Gaia Multilateral Agreement.
The Pan-STARRS1 Surveys (PS1) and the PS1 public science archive have been made possible through contributions by the Institute for Astronomy, the University of Hawaii, the Pan-STARRS Project Office, the Max-Planck Society and its participating institutes, the Max Planck Institute for Astronomy, Heidelberg and the Max Planck Institute for Extraterrestrial Physics, Garching, The Johns Hopkins University, Durham University, the University of Edinburgh, the Queen's University Belfast, the Harvard-Smithsonian Center for Astrophysics, the Las Cumbres Observatory Global Telescope Network Incorporated, the National Central University of Taiwan, the Space Telescope Science Institute, the National Aeronautics and Space Administration under Grant No. NNX08AR22G issued through the Planetary Science Division of the NASA Science Mission Directorate, the National Science Foundation Grant No. AST-1238877, the University of Maryland, Eotvos Lorand University (ELTE), the Los Alamos National Laboratory, and the Gordon and Betty Moore Foundation.
Based on observations obtained at the Hale Telescope, Palomar Observatory, as part of a collaborative agreement between the Caltech Optical Observatories and the Jet Propulsion Laboratory.
The Gordon and Betty Moore Foundation, through both the Data-Driven Investigator Program and a dedicated grant, provided critical funding for SkyPortal.
PS data can be found in MAST \citep{PS_doi}, AllWISE \citep{AllWISE} and Gaia EDR3 \citep{GaiaEDR3} data can be found in IPAC.

\end{acknowledgments}

\vspace{5mm}
\facilities{Zwicky Transient Facility \citep[ZTF, ][]{Bellm_2019, Graham_2019, Masci_2019, Dekany_2020}, Sloan Digital Sky Survey \citep[SDSS, ][]{Almeida_2023}, Pan-STARRS \citep[PS, ][]{Chambers_2016}, Wide-Field Infrared Survey Explorer \citep[WISE, ][]{Wright_2010, Cutri_2014}, Gaia \citep{Gaia_2016, Gaia_2021}}

\software{
Python 3 \citep{python3_2009},
Tensorflow \citep{tensorflow_2015},
Keras \citep{Chollet_2015},
Astromer \citep{Donoso_2023},
XGBoost \citep{Chen_2016},
Scikit-learn \citep{Pedregosa_2011},
Astropy \citep{astropy_2013, astropy_2018, astropy_2022},
SkyPortal \citep{Walt_2019, Coughlin_2023},
NumPy \citep{NumPy_2020},
SciPy \citep{SciPy_2020},
IPython \citep{ipython_2007},
Pandas \citep{pandas_2010},
Matplotlib \citep{Matplotlib_2007},
seaborn \citep{seaborn_2021},
tqdm \citep{tqdm_2019}
}

\appendix

\section{Published files}\label{app:catalog}

\begin{deluxetable}{cc}
\tablenum{6}
\tablecaption{Columns present in the catalog files.}
\tablewidth{0pt}
\tablehead{
\colhead{name} & \colhead{description} \\
}
\decimalcolnumbers
\startdata
ID & ZTF identifier \\
ra & right ascension \\
dec & declination \\
n\_obs & number of ZTF observation epochs \\
is\_duplicate & flag indicating duplicated light curves \\
mag\_median & ZTF \textit{g}-band median magnitude \\
p\_[galaxy, QSO, star] & classification probabilities \\
p\_WISE\_[galaxy, QSO, star] & classifications with added WISE data \\
redshift & redshift estimate \\
\enddata
\tablecomments{Redshift estimate is based on WISE W[1-4] bands, and its presence indicates wether WISE data was is available for a given object. The train data predictions have no ZTF identifiers and duplicate flags.}
\label{tab:columns}
\end{deluxetable}

The \textit{QZO.csv} file includes 4,849,574 objects, and provides columns as described in table \ref{tab:columns}, excluding the duplicate objects flag. The classifications are based on XGB models trained on ZTF \textit{g-band} median magnitude and light curves classification with transformer model, as well as WISE W[1-4] magnitudes and colors. The photo-zs are based on ZTF \textit{g-band} magnitude and WISE magnitudes and colors. We remove duplicated ZTF light curves by removing objects which within the full ZTF catalog have at least one neighbor within $1\arcsec$ with more ZTF observation epochs. The final number of quasars was achieved with magnitude, number of observation epochs, and minimum quasar classification probability cuts, such that $g < n_\mathrm{obs} / 80 + 20.375$, and $p_\mathrm{QSO} > 0.9$, where $p_\mathrm{QSO}$ is XGB classification probability for the quasar class. The photo-zs are available for 35\% of these objects, depending on the availability of WISE observations.

File \textit{ZTF\_all\_QSO.csv} provides all the columns for 78,078,450 objects classified as QSOs by at least one of the two XGB models with and without the WISE features. There are no cuts applied, and there are no duplicates removed. 26\% of objects are marked with the duplicates flag.

We publish the train data predictions in the file \textit{train.csv}. The file contains 2,588,221 records, with ZTF ID and duplicates flag missing. Selecting the longest ZTF light curve for each non duplicated SDSS object removed ZTF duplicates.

\section{Misclassified light curves}\label{app:light_curves}

\begin{figure*}
\fig{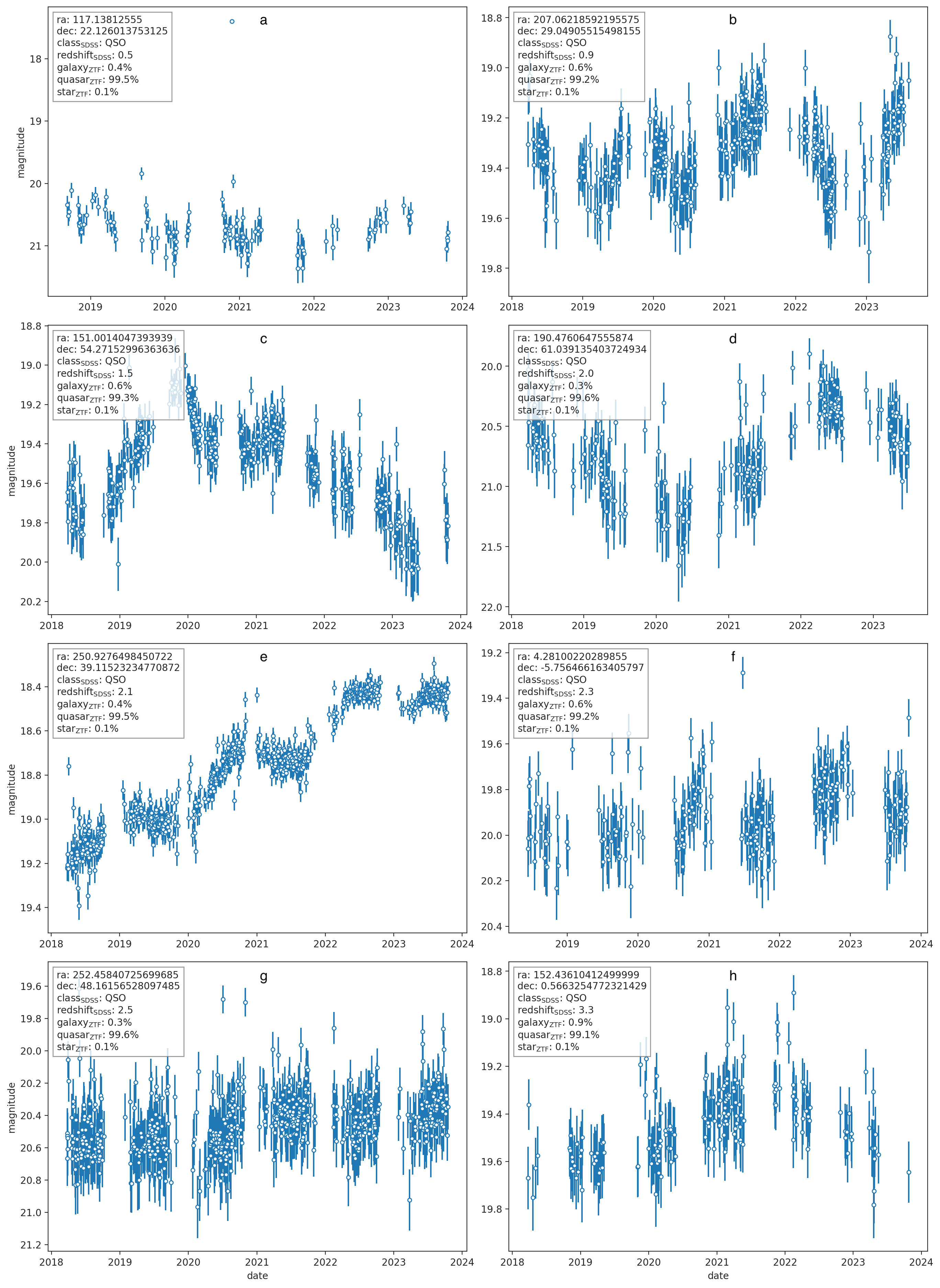}{0.9\textwidth}{}
\caption{Random QSO ZTF \textit{g}-band light curves correctly classified as QSOs by the transformer, sorted by redshift. \label{fig:qso_as_qso}}
\end{figure*}

\begin{figure*}
\fig{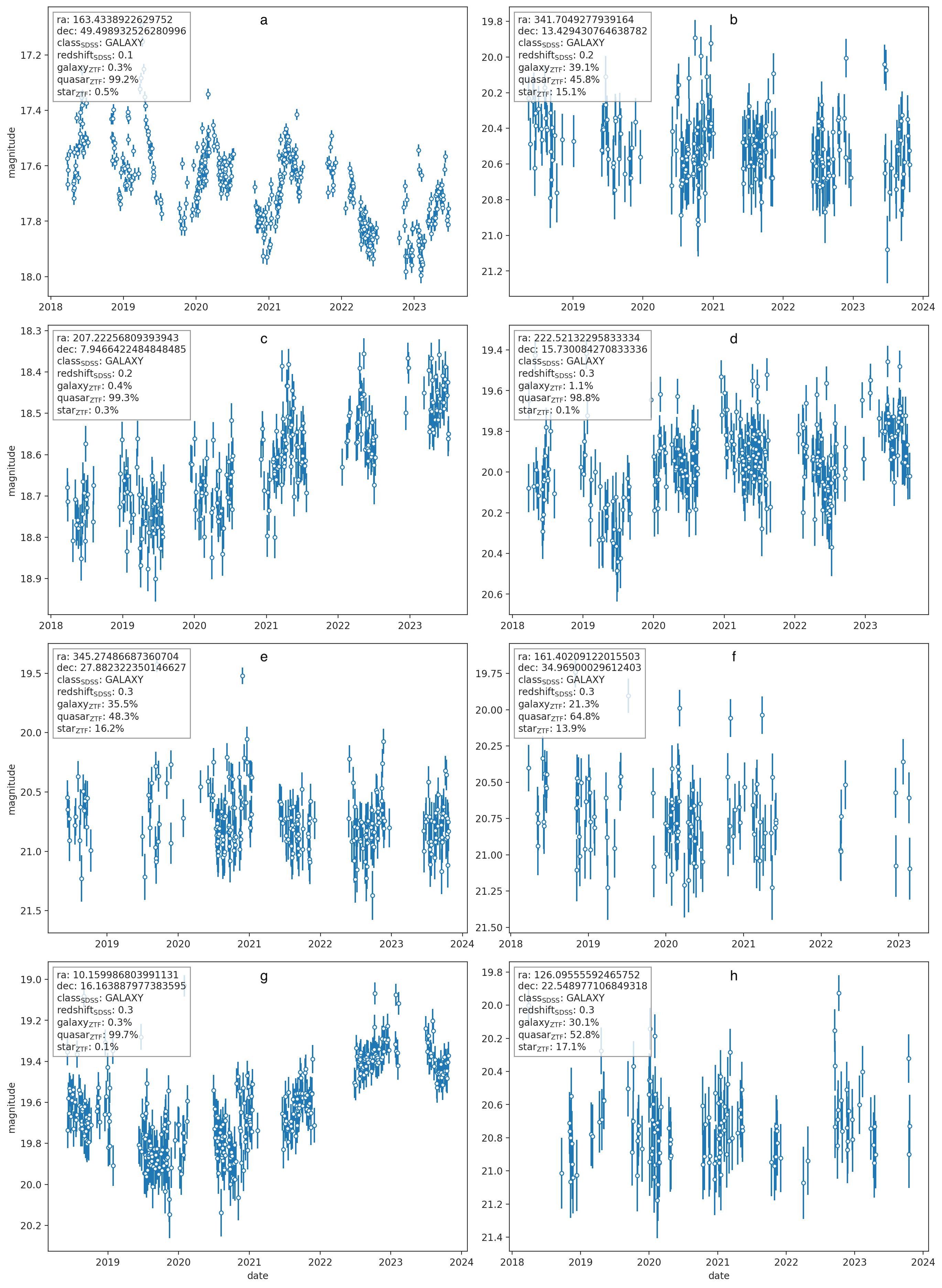}{0.9\textwidth}{}
\caption{Random galaxy ZTF \textit{g}-band light curves classified as QSOs by the transformer, sorted by redshift. \label{fig:galaxy_as_qso}}
\end{figure*}

\begin{figure*}
\fig{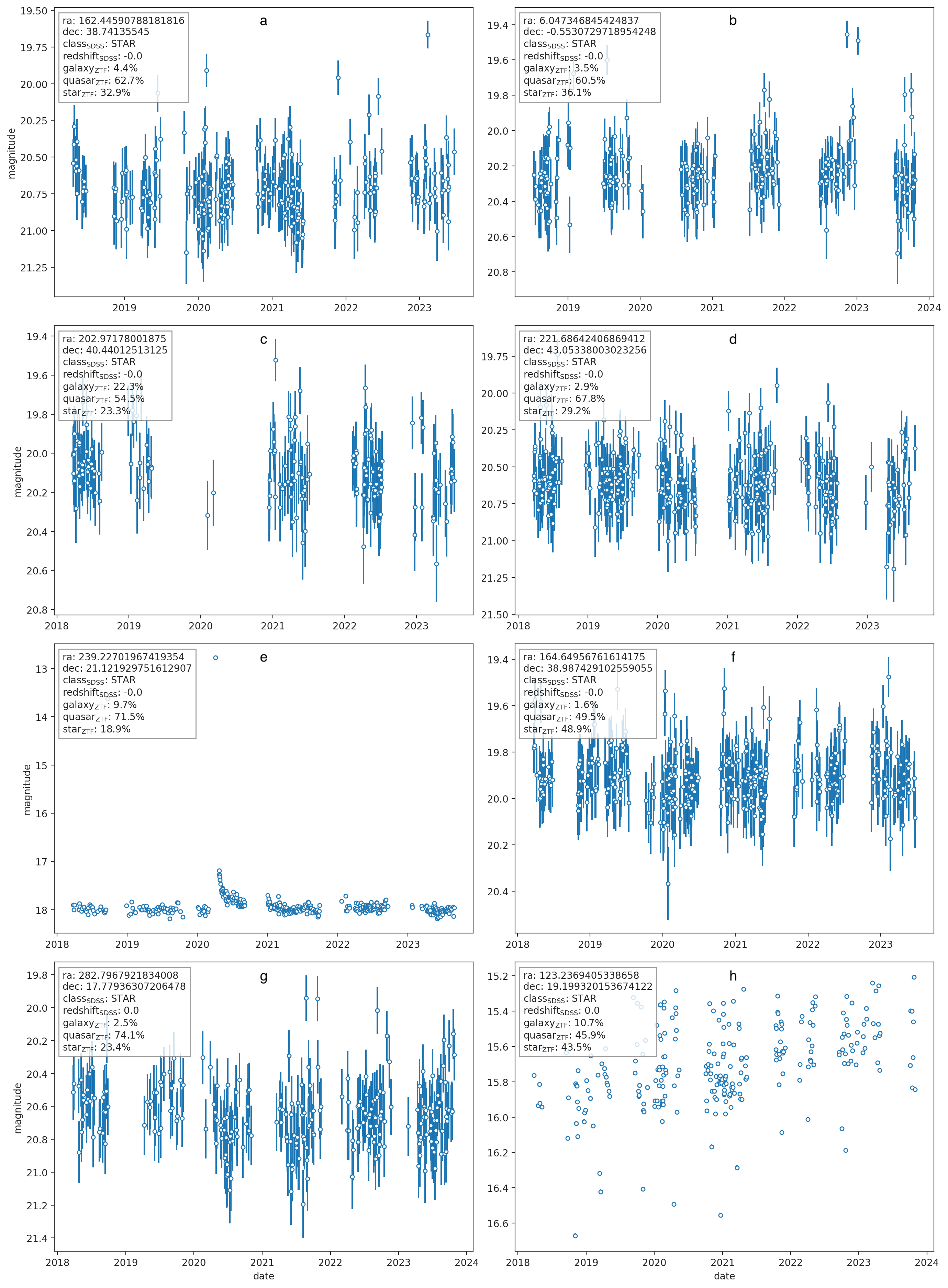}{0.9\textwidth}{}
\caption{Random star ZTF \textit{g}-band light curves incorrectly classified as QSOs by the transformer, sorted by redshift. \label{fig:star_as_qso}}
\end{figure*}

Figures \ref{fig:qso_as_qso}, \ref{fig:galaxy_as_qso} and \ref{fig:star_as_qso} show ZTF \textit{g}-band light curves of QSOs, galaxies and stars, respectively, classified as QSOs by the XGB model based on the transformer classifications and \textit{g}-band median magnitude. This qualitative analysis is based on a small sample of light curves, but shows interesting results. Stars classified as quasars are variable but not periodic, while galaxies classified as quasars can be actually SDSS misclassified AGNs. The QSOs misclassified as galaxies and stars generally show low amplitude variability.

\bibliography{main}{}
\bibliographystyle{aasjournal}

\end{document}